\documentclass[sigconf]{acmart}

\usepackage{todonotes}
\usepackage{xspace}
\usepackage{multicol}

\usepackage{listings}
\usepackage{listingsLanguages}
\usepackage{enumitem}
\definecolor{maroon}{cmyk}{0, 0.87, 0.68, 0.32}
\definecolor{halfgray}{gray}{0.55}
\definecolor{ipython_frame}{RGB}{207, 207, 207}
\definecolor{ipython_bg}{RGB}{247, 247, 247}
\definecolor{ipython_red}{RGB}{186, 33, 33}
\definecolor{ipython_green}{RGB}{0, 128, 0}
\definecolor{ipython_blue}{RGB}{64, 128, 128}
\definecolor{ipython_purple}{RGB}{170, 34, 255}
\usepackage{xcolor,colortbl}
\usepackage[frozencache=true,cachedir=.]{minted}
\usepackage{bbding}

\newcommand{\sys}{\texttt{ir\_datasets}\xspace}

\newcommand{\chk}{\CheckmarkBold}
\newcommand{\loz}{$\lozenge$}
\newcommand{\out}{\rightarrowtail}

\newcommand{\p}[1]{\phantom{#1}}
\newcommand{\g}[1]{{\color{gray}#1}}
\newcommand{\f}[1]{\textbf{#1}}

\AtBeginDocument{%
  \providecommand\BibTeX{{%
    \normalfont B\kern-0.5em{\scshape i\kern-0.25em b}\kern-0.8em\TeX}}}

\copyrightyear{2021} 
\acmYear{2021} 
\setcopyright{acmcopyright}
\acmConference[SIGIR '21]{Proceedings of the 44th International ACM SIGIR Conference on Research and Development in Information Retrieval}{July 11--15, 2021}{Virtual Event, Canada}
\acmBooktitle{Proceedings of the 44th International ACM SIGIR Conference on Research and Development in Information Retrieval (SIGIR '21), July 11--15, 2021, Virtual Event, Canada}
\acmPrice{15.00}
\acmDOI{10.1145/3404835.3463254}
\acmISBN{978-1-4503-8037-9/21/07}

\begin{document}

\title{Simplified Data Wrangling with \sys}

\author{Sean MacAvaney}\authornote{This work was conducted in part during an internship at the Allen Institute for AI.}
\affiliation{\institution{University of Glasgow}\country{}}
\affiliation{\institution{IR Lab, Georgetown University}\country{}}
\email{sean.macavaney@glasgow.ac.uk}

\author{Andrew Yates}
\affiliation{\institution{Max Planck Institute for Informatics}\country{}}
\email{ayates@mpi-inf.mpg.de}

\author{Sergey Feldman, Doug Downey, Arman Cohan}
\affiliation{\institution{Allen Institute for AI}\country{}}
\email{{sergey,dougd,armanc}@allenai.org}

\author{Nazli Goharian}
\affiliation{\institution{IR Lab, Georgetown University}\country{}}
\email{nazli@ir.cs.georgetown.edu}

\renewcommand{\shortauthors}{MacAvaney, et al.}




\begin{abstract}
Managing the data for Information Retrieval (IR) experiments can be challenging. Dataset documentation is scattered across the Internet and once one obtains a copy of the data, there are numerous different data formats to work with. Even basic formats can have subtle dataset-specific nuances that need to be considered for proper use.
To help mitigate these challenges, we introduce a new robust and lightweight tool (\sys) for acquiring, managing, and performing typical operations over datasets used in IR. We primarily focus on textual datasets used for ad-hoc search. This tool provides both a Python and command line interface to numerous IR datasets and benchmarks. To our knowledge, this is the most extensive tool of its kind. Integrations with popular IR indexing and experimentation toolkits demonstrate the tool's utility. We also provide documentation of these datasets through the \sys catalog: \url{https://ir-datasets.com/}. The catalog acts as a hub for information on datasets used in IR, providing core information about what data each benchmark provides as well as links to more detailed information. We welcome community contributions and intend to continue to maintain and grow this tool.
\end{abstract}




\maketitle

\section{Introduction}

The datasets and benchmarks we use are a cornerstone of Information Retrieval (IR) research. Unfortunately, many of these datasets remain frustrating to find and manage. Once obtained, the variety of data formats can be a challenge to work with. 
Even data formats that seem simple can hide subtle problems. For example, the TSV files used by the MS-MARCO~\cite{Bajaj2016MSMA} has a double-encoding problem that affects special characters in roughly 20\% of documents.

Recently, several tools have begun to incorporate automatic dataset acquisition. These include Capreolus~\cite{yates2020capreolus}, PyTerrier~\cite{pyterrier2020ictir} and OpenNIR~\cite{macavaney:wsdm2020-onir}. These reduce the user burden of finding the dataset source files and figuring out how to parse them correctly. However, the dataset coverage of each individually is patchy, as shown in Table~\ref{tab:dataset_comp}. Further, using the dataset interfaces outside of these tools can be difficult, as they are often tightly coupled with the tool's primary functionality. Finally, each of these tools keep their own copy of data, leading to wasted storage. Thus, it is advantageous to have a lightweight tool that focuses on data acquisition, management, and typical operations like lookups.

Many tools rely on manual instructions for downloading, extracting, and processing datasets.\footnote{\label{ftn:msm}Such as \url{https://github.com/castorini/anserini/blob/master/docs/experiments-msmarco-passage.md}, \url{https://github.com/thunlp/OpenMatch/blob/master/docs/experiments-msmarco.md}, \url{https://github.com/microsoft/ANCE\#data-download}, etc.}
We believe providing a tool to automatically perform as much of this work as possible is clearly preferable to this approach since it ensures proper processing of data.
A \textit{common} automatic tool has additional advantages, such as reducing redundant copies of datasets and easily allowing tools to be run on alternative or custom datasets with little effort.

\begin{table}
\centering
\caption{Dataset support in Capreolus~\cite{yates2020capreolus} (Cap.), PyTerrier~\cite{pyterrier2020ictir} (PT), OpenNIR~\cite{macavaney:wsdm2020-onir} (ONIR), Anserini~\cite{Yang2017AnseriniET} (Ans.), and \sys (IRDS). \chk~indicates built-in support that automatically provides documents, queries, and query relevance judgments (i.e., as an automatic download). \loz~indicates support for a dataset with some manual effort (e.g., specifying the document parser and settings to use). Datasets marked with * have licenses that require manual effort (e.g., requesting from NIST), and therefore can at most have \loz.
}
\label{tab:dataset_comp}
\vspace{-0.5em}
\begin{tabular}{lccccc}
\toprule
Dataset/Benchmark                                 &Cap.& PT &ONIR&Ans.&IRDS\\
\midrule
\bf News \\
NYT*~\cite{sandhaus2008nyt,macavaney:nyt}          &    &\loz&\loz&\loz&\loz\\
TREC Arabic*~\cite{LDC2001T55,arabic01,arabic02}   &    &\loz&\loz&\loz&\loz\\
TREC Common Core*~\cite{allan2017treccore}         &    &\loz&    &\loz&\loz\\
TREC Mandarin*~\cite{LDC2000T52,zh5,zh6}           &    &\loz&\loz&\loz&\loz\\
TREC News*~\cite{news18,news19}                    &    &\loz&    &\loz&\loz\\
TREC Robust*~\cite{Voorhees2004Rb,Voorhees2005Rb}  &\loz&\loz&\loz&\loz&\loz\\
TREC Spanish*~\cite{LDC2000T51,es3,es4}            &    &\loz&\loz&\loz&\loz\\
\midrule\bf Question Answering \\
ANTIQUE~\cite{Hashemi2020Antique}                 &\chk&\loz&\chk&\loz&\chk\\
MS-MARCO Doc.~\cite{Bajaj2016MSMA}                &    &\chk&\loz&\chk&\chk\\
MS-MARCO Pass.~\cite{Bajaj2016MSMA}               &\chk&\chk&\chk&\chk&\chk\\
MS-MARCO QnA~\cite{Bajaj2016MSMA}                 &    &    &    &    &\chk\\
Natural Questions~\cite{nq,dprnq}                 &    &    &    &\loz&\chk\\
TREC CAR~\cite{Dietz2017,dietz2017trec}           &    &    &\chk&\loz&\chk\\
TREC DL~\cite{dl19,dl20}                          &\chk&\chk&\chk&\chk&\chk\\
TREC DL-Hard~\cite{Mackie2021-howdeep}            &    &\loz&\loz&\loz&\chk\\
TriviaQA~\cite{Joshi2017TriviaQAAL,dprnq}         &    &    &    &\loz&\chk\\
\midrule\multicolumn{5}{l}{\bf Scientific, Bio-medical, Health} \\
Cranfield~\cite{cranfield}                        &    &    &    &    &\chk\\
CLEF eHealth*~\cite{Zuccon2016CLEF,Palotti2017CLEF}&    &\loz&    &\loz&\loz\\
NFCorpus~\cite{boteva16full}                      &\chk&    &    &    &\chk\\
TREC CDS~\cite{cds2014,cds2015,cds2016}           &    &    &    &    &\chk\\
TREC COVID~\cite{Wang2020CORD19TC,treccovid}      &\chk&\chk&\chk&\chk&\chk\\
TREC Genomics~\cite{gen04,gen05,gen06,gen07}      &    &    &    &    &\chk\\
TREC Health Misinfo.*~\cite{trecdecision2019}      &    &\loz&    &\loz&\loz\\
TREC PM~\cite{pm17,pm18,pm19}                     &    &\loz&    &    &\chk\\
TripClick*~\cite{rekabsaz2021tripclick}            &    &\loz&    &\loz&\loz\\
Vaswani~\cite{vaswani}                            &    &\chk&    &    &\chk\\
\midrule\bf Web \\
NTCIR WWW*~\cite{ntcirwww1,ntcirwww2}              &    &\loz&    &\loz&\loz\\
ORCAS~\cite{craswell2020orcas}                    &    &\loz&\loz&\loz&\chk\\
\multicolumn{2}{l}{TREC Million Query*~\cite{mq07,mq08,mq09}}&\loz&    &\loz&\loz\\
TREC Terabyte*~\cite{tb04,tb05,tb06}               &    &\loz&    &\loz&\loz\\
\multicolumn{2}{l}{TREC Web*~\cite{w02,w03,w04,w09,w10,w11,w12,w13,w14}}&\loz&    &\loz&\loz\\
\midrule\multicolumn{5}{l}{\bf Other/Miscellaneous} \\
\multicolumn{5}{l}{BEIR~\cite{beir,ag,cf,dbp,fv,fi,hp,Bajaj2016MSMA,boteva16full,nq,sd,sf,Wang2020CORD19TC,treccovid,to,cqa}}&\chk\\
CodeSearchNet~\cite{Husain2019CodeSearchNet}      &\chk&    &    &    &\chk\\
TREC Microblog~\cite{Sequiera2017Finally,mb13,mb14}&   &    &    &\loz&\chk\\
WikIR~\cite{wikir,mlwikir}                        &    &    &\chk&    &\chk\\
\bottomrule
\end{tabular}
\end{table}

Anserini~\cite{Yang2017AnseriniET} and its Python interface Pyserini~\cite{Lin2021PyseriniAE} use a hybrid approach by distributing copies of queries and relevance judgments in the package itself and primarily relying on manual instructions for document processing. Sometimes Anserini provides document content via downloadable indices.

Other dataset distribution tools are not well-suited for IR tasks. For instance, packages like HuggingFace Datasets~\cite{2020HuggingFace-datasets} and TensorFlow Datasets~\cite{TFDS} take a record-centric approach that is not well-suited for relational data like documents, queries, and query-document relevance assessments. Furthermore, IR work involves additional important use cases when working with datasets, such as efficiently looking up a document by ID, for which the designs of prior libraries is not conducive.
Dataset schemata, such as DCAT and schema.org, provide a common format machine-readable dataset documentation, which could be supported in the future.

\begin{table*}
\centering
\caption{Entity types in \sys.}
\label{tab:entites}
\vspace{-1em}
\begin{tabular}{llp{4.7in}}
\toprule
Entity Type & Python API Example & Description \\
\midrule
docs & \texttt{ds.docs\_iter()} & A document (or passage for passage retrieval). Contains a \texttt{doc\_id} and one or more text fields. \\
queries & \texttt{ds.queries\_iter()} & A query (topic). Contains a \texttt{query\_id} and one or more text fields. \\
qrels & \texttt{ds.qrels\_iter()} & A query relevance assessment. Maps a \texttt{query\_id} and \texttt{doc\_id} to a relevance score or other human assessments. \\
scoreddocs & \texttt{ds.scoreddocs\_iter()} & (uncommon) A scored document (akin to a line from a run file). Maps a \texttt{query\_id} and \texttt{doc\_id} to a ranking score from a system. Available for datasets that provide an initial ranking (for testing reranking systems). \\
docpairs & \texttt{ds.docpairs\_iter()} & (uncommon) A pair of documents (useful for training). Maps a \texttt{query\_id} to two or more \texttt{doc\_id}s. Available for datasets that provide suggested training pairs. \\
\bottomrule
\end{tabular}
\end{table*}

In this work, we present \sys, a tool to aid IR researchers in the discovery, acquisition, and management of a variety of IR datasets. The tool provides a simple and lightweight Python and command line interface (see Figure~\ref{fig:primary}) allowing users to iterate the documents, queries, relevance assessments, and other relations provided by a dataset. This is useful for indexing, retrieval, and evaluation of ad-hoc retrieval systems. A document lookup API provides fast access to source documents, which is useful for recent text-based ranking models, such as those that use BERT~\cite{DevlinCLT19}. PyTerrier~\cite{pyterrier2020ictir}, Capreolus~\cite{yates2020capreolus}, and OpenNIR~\cite{macavaney:wsdm2020-onir} recently added support for \sys, greatly expanding the number of datasets they support, and other tools like Anserini~\cite{Yang2017AnseriniET} can utilize our tool using the command line interface. Finally, the \sys catalog\footnote{\label{ftn:catalog}\url{https://ir-datasets.com/}} acts as a documentation hub, making it easy to find datasets and learn about their characteristics. We intend to continue to backfill prior datasets and add support for new datasets as they are released. The package is open source,\footnote{\url{https://github.com/allenai/ir_datasets/}} and we welcome contributions.

\section{\sys}

\sys is a lightweight tool focused on providing easy access to a variety of IR datasets and benchmarks. It provides both a Python and command line interface (see Figure~\ref{fig:primary}), allowing it to be easily used by a variety of toolkits, or simply for ad-hoc data exploration.

\begin{figure}

\begin{lstlisting}[
    linewidth=\columnwidth,
    escapeinside={(*}{*)},
    language=Python]
import ir_datasets
dataset = ir_datasets.load('msmarco-passage/train')
for doc in datasets.docs_iter(): # documents
    print(doc)
(*$\out$*) # GenericDoc(doc_id='0', text='The presence of commun...
(*$\out$*) # GenericDoc(doc_id='1', text='The Manhattan Project ...
(*$\out$*) # ...

for query in dataset.queries_iter(): # queries
    print(query)
(*$\out$*) # GenericQuery(query_id='121352', text='define extreme')
(*$\out$*) # GenericQuery(query_id='634306', text='what does chatt...
(*$\out$*) # ...

for qrel in dataset.qrels_iter(): # relevance judgments
    print(qrels)
(*$\out$*) # TrecQrel(query_id='1185869', doc_id='0', relevance=1)
(*$\out$*) # TrecQrel(query_id='1185868', doc_id='16', relevance=1)
(*$\out$*) # ...

# Look up documents by ID
docs_store = dataset.docs_store()
docs_store.get("16")
(*$\out$*) # GenericDoc(doc_id='16', text='The approach is based...
\end{lstlisting}

\begin{lstlisting}[
    linewidth=\columnwidth,
    escapeinside={(*}{*)},
    language=bash]
$ ir_datasets export msmarco-passage docs
(*$\out$*) # 0	The presence of communication amid scientific...
(*$\out$*) # 1	The Manhattan Project and its atomic bomb hel...
(*$\out$*) # ...

$ ir_datasets export msmarco-passage/train queries
(*$\out$*) # 121352	define extreme
(*$\out$*) # 634306	what does chattel mean on credit history
(*$\out$*) # ...

$ ir_datasets export msmarco-passage/train qrels
(*$\out$*) # 1185869 0 0 1
(*$\out$*) # 1185868 0 16 1
(*$\out$*) # ...

# Look up documents by ID
$ ir_datasets lookup msmarco-passage/train 16
(*$\out$*) # 16	The approach is based on a theory of justice...
\end{lstlisting}

\caption{Parallel examples of common use cases in \sys using Python and the command line interface.}\label{fig:primary}

\end{figure}

To achieve these goals, \sys adheres to several design principles. First, to stay lightweight, the tool is focused on core dataset operations, such as downloading content, iterating through queries or documents, and performing document lookups by ID. This policy explicitly leaves functionality like full-text indexing or neural network processing to other tools. Further, to be practical in a variety of environments, \sys attempts to keep a low memory footprint by using inexpensive data structures and iterators. Finally, in order to leave maximum flexibility to the tool's users, we attempt to perform ``just enough'' processing of the data to account for various formats, while not removing information that is potentially useful. We hope that this commitment to being lightweight and flexible makes \sys an attractive tool to jump-start or enhance other tools for doing IR research.

\subsection{Dataset Identifiers}

Since no standard identifiers (IDs) exist for datasets in IR, we propose hierarchical dataset IDs. These IDs allow datasets to be looked up in the Python API, command line interface, and online documentation. IDs are usually in the format of \texttt{corpus/benchmark}. For instance, the TREC COVID~\cite{treccovid} benchmark uses the CORD-19~\cite{Wang2020CORD19TC} document corpus and is given an ID of \texttt{cord19/trec-covid}. In this case, \texttt{cord19} provides documents, while \texttt{cord19/trec-covid} provides queries and relevance judgments for those documents.

\subsection{Simple \& Memorable Python API}

A dataset object can be obtained simply by calling:
\begin{minted}[fontsize=\small]{python}
import ir_datasets
ds = ir_datasets.load("dataset-id")
\end{minted}
Each dataset objects provides access to a number of entity types (see Table~\ref{tab:entites}). Dataset objects are stateless; they simply define the capabilities and the procedures for obtaining and processing the data.

Most ad-hoc retrieval datasets consist of 3 main entity types: documents (\texttt{docs}), queries/topics (\texttt{queries}), and query relevance assessments (\texttt{qrels}).
In the spirit of being simple, lightweight, and low-memory, entities are provided as \texttt{namedtuple} instances from iterators. For each entity type provided by a particular dataset, there is a corresponding \texttt{ds.\{entity\}\_iter()} function that returns an iterator (e.g., \texttt{ds.docs\_iter()}). Since the particular attributes returned for an entity differ between datasets (e.g., some provide only an ID and text for a document, while others also include a title field), type definitions can be accessed via \texttt{ds.\{entity\}\_cls()}. The type definitions include type annotations for each field, and try to adhere to conventions when possible (e.g., the ID of documents is the first field and named \texttt{doc\_id}).

The iterator approach is versatile. In some cases, it is only necessary to operate over a single entity at a time, minimizing the memory overhead. In other cases, particularly in neural networks, operations happen in batches, which can also be accomplished trivially through an iterator. And finally, in cases where all data needs to be loaded, all entities can be easily loaded, e.g., by passing the iterator into the Python list constructor, or the dataframe constructor in Pandas~\cite{reback2020pandas}.

Some datasets provide other entity types, such as sample document rankings or training sequences. For the former, we have a \texttt{scoreddocs} entity type, which by default is a tuple containing a query ID, a document ID, and a score. For the latter, we have a \texttt{docpairs} entity, which consists of a query and a pair of contrasting document IDs (e.g., one relevant and one non-relevant).

\subsection{Command Line Interface}

\sys also provides a Command Line Interface (CLI) for performing basic operations over supported datasets. This is helpful for integration with tools not written in Python, or simply for ad-hoc data exploration. The primary operations of the CLI are \texttt{export} (corresponding to Python's \texttt{dataset.*\_iter()} functions) and \texttt{lookup} (corresponding to Python's \texttt{docstore.get\_many\_iter()}). Examples of these operations are shown in right-hand side of Figure~\ref{fig:primary}. The command line interface supports multiple output formats, including TSV and JSON lines. The output fields can also be specified, if only certain data is desired.

\subsection{Data Acquisition}

When possible, \sys downloads content automatically from the original public sources as needed. In cases where a data usage agreement exists, the user is notified before the file is downloaded. The download process is robust; it verifies the integrity of the downloaded content via a hash and is resilient to interrupted downloads by re-issuing the request if the connection is broken (using Range HTTP requests, if supported by the server). Further, the access to and integrity of downloadable content is automatically checked periodically using a continuous integration job so that if access to some resources are lost (e.g., a file is moved) the problem can be quickly investigated and fixed. There are nearly 350 downloadable files supporting the current datasets in \sys, each validated weekly.

Some data are not publicly available. For instance, due to its size, the ClueWeb 2009 and 2012 collections (used for tasks like the TREC WebTrack and NTCIR WWW tasks) are obtained via hard drives. Other datasets, like the Arabic Newswire collection (used for the TREC Arabic tasks) contain copyrighted material and are only available with a usage agreement and subscription to the Linguistic Data Consortium. In these cases, the user is presented with instructions on how to acquire the dataset and where to put it. Once acquired by the user, \sys will take care of any remaining processing. There are currently 12 document collections that require a manual process to acquire.

\subsection{Supported datasets}

\sys supports a wide variety of datasets (see Table~\ref{tab:dataset_comp}). These include some of the most popular evaluation benchmarks (e.g., TREC Robust~\cite{Voorhees2004Rb}), large-scale shallow datasets (e.g., MS-MARCO~\cite{Bajaj2016MSMA}), biomedical datasets (e.g., TREC CDS~\cite{cds2014,cds2015,cds2016}), multi- and cross-lingual datasets (e.g., TREC Arabic~\cite{arabic01,arabic02}), a content-based weak supervision dataset (NYT~\cite{macavaney:nyt}), a large-scale click dataset (ORCAS~\cite{craswell2020orcas}), and a ranking benchmark suite (BEIR~\cite{beir}). To our knowledge, this represents the largest collection and variety of IR datasets supported by any tool. To facilitate experiments with custom datasets, the Python API provides an easy mechanism to build a dataset object from files that use simple data formats:
\begin{minted}[fontsize=\small]{python}
ds = ir_datasets.create_dataset(docs_tsv="path/docs.tsv",   
  queries_tsv="path/queries.tsv", qrels_trec="path/qrels")
\end{minted}

\subsection{Document lookups}

It is a common task to look up documents by their ID. For instance, when training or running a neural IR model, it is often necessary to fetch the text of the current document to perform processing. Another example would be a researcher who is looking into cases in which their model fails may want to see the text of the offending documents.

One option is to load all documents into an in-memory hashmap. This may be appropriate in some cases, such a long-running process where the large upfront cost is negligible and memory is plentiful (enough for the entire collection). Building an in-memory hashmap for a collection is trivial with the Python interface:
\begin{minted}[fontsize=\small]{python}
doc_map = {doc.doc_id: doc for doc in dataset.docs_iter()}
\end{minted}

To support other cases, \sys provides a \texttt{docs\_store} API that simplifies the process of looking up documents from disk. This API supports fetching individual or multiple documents by their ID:

\begin{minted}[fontsize=\small]{python}
docs_store = dataset.docs_store()
docs_store.get_many(['D1', 'D2'])
# {'D1': GenericDoc('D1', ...), 'D2': GenericDoc('D2', ...)}
it = docs_store.get_many_iter(['D1', 'D2'])
# An iterator of D1 and D2 (order not guaranteed)
\end{minted}

\begin{table}
\centering
\caption{Document lookup benchmarks on small datasets.}
\label{tab:doc_lookup}
\vspace{-1em}
\begin{tabular}{lrrrr}
\toprule
& \multicolumn{3}{c}{Time/query} \\
\cmidrule(lr){2-4}
System& HDD\p{ s} & SSD\p{ ms} & Warm\p{ ms}  & Size\p{ GB} \\
\midrule
\multicolumn{5}{c}{\texttt{msmarco-passage/trec-dl-2019} (avg. 949 docs/query)} \\
\midrule
\sys     &\f{2.34}\g{~s} &\f{66}\g{~ms} &\f{7}\g{~ms} &   {2.8}\g{~GB} \\
MongoDB  &  {3.62}\g{~s} & {130}\g{~ms} & {14}\g{~ms} &   {2.7}\g{~GB} \\
SQLite   &  {3.72}\g{~s} &  {94}\g{~ms} & {27}\g{~ms} &   {4.1}\g{~GB} \\
Pyserini &\f{2.34}\g{~s} &  {85}\g{~ms} & {51}\g{~ms} &   {2.4}\g{~GB} \\
PyTerrier&  {3.40}\g{~s} & {138}\g{~ms} & {68}\g{~ms} & \f{2.3}\g{~GB} \\
\midrule
\multicolumn{5}{c}{\texttt{cord19/fulltext/trec-covid} (avg. 1,386 docs/query)} \\
\midrule
\sys     &\f{1.19}\g{~s} &\f{0.11}\g{~s}\p{m} & \f{36}\g{~ms} & \f{1.3}\g{~GB} \\
MongoDB  &  {3.65}\g{~s} &  {0.19}\g{~s}\p{m} &   {65}\g{~ms} &   {1.8}\g{~GB} \\
SQLite   &  {5.99}\g{~s} &  {0.19}\g{~s}\p{m} &   {50}\g{~ms} &   {2.8}\g{~GB} \\
Pyserini &  {2.05}\g{~s} &  {0.32}\g{~s}\p{m} &   {51}\g{~ms} &   {1.5}\g{~GB} \\
PyTerrier&  {3.72}\g{~s} &  {1.70}\g{~s}\p{m} &{1,620}\g{~ms} &   {4.2}\g{~GB} \\
\bottomrule
\end{tabular}
\end{table}

\begin{table}
\centering
\caption{Document lookup benchmarks on large datasets. Storage costs are listed as space beyond the source files.}
\label{tab:doc_lookup_cw12}
\vspace{-1em}
\begin{tabular}{lrrr}
\toprule
& \multicolumn{2}{c}{Time/query} \\
\cmidrule(lr){2-3}
Strategy& HDD\p{ s} &Warm\p{ ms}  & Size\p{ GB} \\
\midrule
\multicolumn{4}{c}{\texttt{clueweb12/trec-web-2014} (avg. 289 docs/query)} \\
\midrule
\sys                         & {44.4}\g{~s} &  \f{14}\g{~ms} &    {4.5}\g{~GB} \\
\hspace{1em}(w/o checkpoints)&{369.4}\g{~s} &  \f{14}\g{~ms} &  \f{0.3}\g{~GB} \\
Pyserini  &\f{19.7}\g{~s} &{1,210}\g{~ms} &{6,041.5}\g{~GB} \\

\midrule
\multicolumn{4}{c}{\texttt{tweets2013-ia/trec-mb-2013} (avg. 1,188 docs/query)} \\
\midrule
\sys                         & {23.3}\g{~s} &  \f{24}\g{~ms} &\f{120}\g{~GB} \\
Pyserini                    &\f{17.6}\g{~s} &   {115}\g{~ms} &  {323}\g{~GB} \\

\bottomrule
\end{tabular}
\end{table}

The implementation of \texttt{docs\_store()} varies based on the dataset. For many small datasets (those with up to a few million documents), we build a specialized lookup structure for the entire collection on disk as needed. A specialized structure was built for this package to provide a good trade-off between lookup speed and storage costs. All documents are compressed using lz4
and stored in sequence. A separate sorted document ID and corresponding index offset structure is also built on disk. Although simple, we found that this structure enables lookups that exceed the performance of leading indexes and databases (see Table~\ref{tab:doc_lookup}). In this experiment, we used the metadata lookup functionality of Anserini~\cite{Yang2017AnseriniET} and Terrier~\cite{ounis2005terrier} and key-value storage with SQLite and MongoDB. The average duration was computed per query for TREC DL 2019 passage task~\cite{dl19} (with the official set of reranking documents), and for TREC COVID complete~\cite{treccovid} (using the judged documents). We also find that the storage cost is reasonable, with a total storage size comparable to MongoDB for the MS-MARCO passage collection and smaller than all others for the CORD19 collection.

For large collections, it is impractical and undesirable to make a copy of all documents. For instance, the ClueWeb09 and ClueWeb12 collections (for TREC Web Track) are several TB in size, even when heavily compressed. Luckily, for these datasets, their directory structure mimics the structure of the document IDs, which allows the desired source file containing a given document ID to be easily identified. To speed up lookups within a given file, we use \texttt{zlib-state}~\footnote{\url{https://github.com/seanmacavaney/zlib-state}} to take periodic checkpoints of the zlib decoding state of the source files. This eliminates the need to read all the source file contents up to the desired document and greatly speeds up lookups of documents that appear late in the source files. The pre-built checkpoints are automatically downloaded and used when appropriate. Furthermore, we cache fetched documents on disk for even faster subsequent lookups. Different approaches are taken for other large collections, such as Tweets2013-ia~\cite{Sequiera2017Finally} (for the TREC Microblog task~\cite{mb13,mb14}). See Table~\ref{tab:doc_lookup_cw12} for a comparison between document lookup times using \sys and Pyserini (from stored document source). Even though \sys is slower than Pyserini on the first lookup, the cache greatly speeds up subsequent fetches (see ``Warm''). Since experiments in neural IR frequently only work with a small subset of documents, this is very beneficial for these pipelines. We also observe that the checkpoint files for ClueWeb12 speed up lookups considerably, without adding much overhead in terms of storage; since Anserini keeps a copy of all documents, it accumulates around 6TB of storage overhead, compared to 4.5GB using \sys. Note that the other approaches explored in Table~\ref{tab:dataset_comp} would accumulate similar storage overheads, as they also copy the data. Tweets2013-ia accumulates considerable storage costs, as the source hierarchy is not conducive to document lookups. In this case, \sys builds an ID-based lookup file hierarchy.

\subsection{Fancy slicing}

In many cases, it is beneficial to select a segment of a document collection. For instance, some techniques involve pre-computing neural document representations to speed up reranking~\cite{macavaney:sigir2020-eff} or for performing first-stage retrieval~\cite{Khattab2020ColBERTEA}. In this case, dividing the operation over multiple GPUs or machines can yield substantial speed gains, as the process is embarrassingly parallel. To divide up the work, it is helpful to be able to select ranges of the document collection for processing.

The Python standard library \texttt{islice} function is not ideal for this task because I/O and processing of documents would be performed for skipped indices. Instead, all objects returned form \texttt{doc\_iter} can themselves be sliced directly. The implementation of the slicing depends on the particular dataset, but in all implementations avoid unnecessary I/O and processing by seeking to the appropriate location in the source file. This \textit{fancy slicing} implementation mostly follows typical Python slicing semantics, allow for different workers to be assigned specific ranges of documents:

\begin{minted}[fontsize=\small]{python}
dataset.docs_iter()[:10] # the first 10 docs
dataset.docs_iter()[-10:] # the last 10 docs
dataset.docs_iter()[100:110] # 10 docs starting at index 100
dataset.docs_iter()[3::5] # every 5 docs, starting at index 3
dataset.docs_iter()[:1/3] # the first third of the collection
\end{minted}

\subsection{Documentation}

\begin{figure}
\centering
\fbox{\includegraphics[scale=0.4]{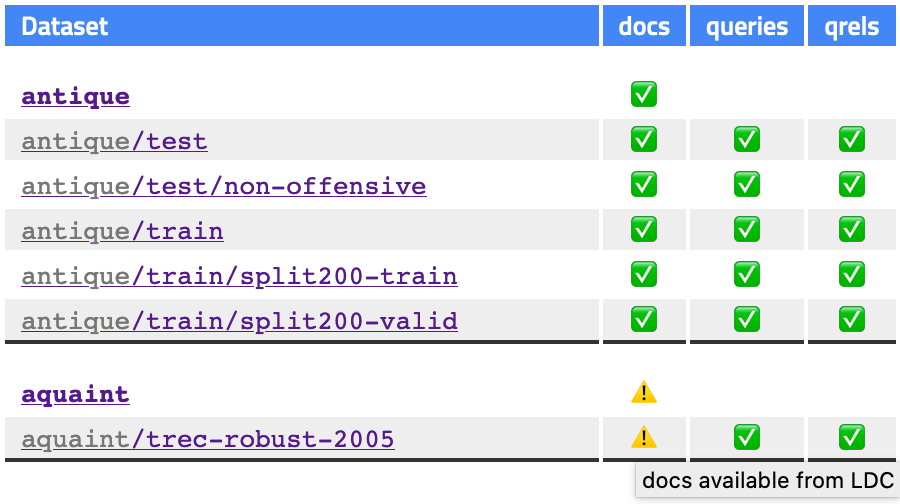}}
\vspace{-0.5em}
\caption{Example from the \sys catalog. Users can easily check which datasets are available for automatic downloads (green checkbox) and those that require obtaining data from a third party (yellow triangle).}
\label{fig:catalog-index}
\end{figure}

\begin{figure}
\centering
\fbox{\includegraphics[scale=0.4]{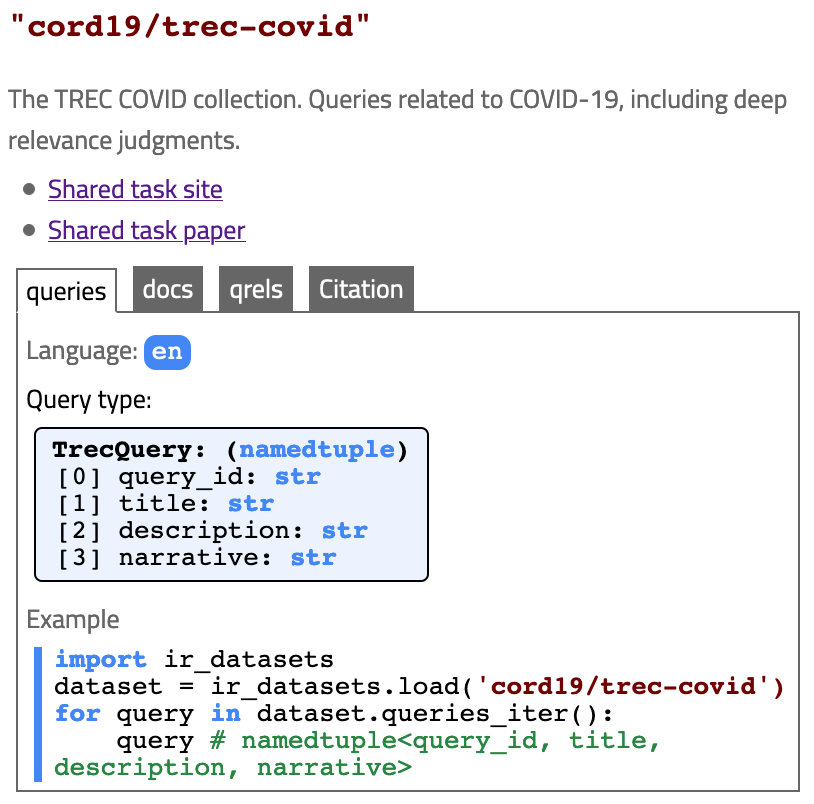}}
\vspace{-0.5em}
\caption{Example documentation for \texttt{cord19/trec-covid}.}
\label{fig:docs}
\end{figure}

Documentation about datasets are available from the \sys catalog.\footnotemark[4] An overview list shows all available datasets and their capabilities (Figure~\ref{fig:catalog-index}). The documentation page for each individual dataset includes a brief description, relevant links (e.g., to shared task website and paper), supported relations, citations, and code samples. An example is shown in Figure~\ref{fig:docs} for the TREC COVID dataset~\cite{treccovid}.

\subsection{Automated Testing}


\sys includes several suites of automated tests to ensure the package works as expected, functionality does not regress as changes are made, and to ensure that downloaded content remains available and unchanged. The automated testing suite includes include unit tests, integration/regression tests, and tests to ensure downloadable content remains available and unchanged.




\section{Integration with Other Tools}\label{sec:others}

The CLI makes \sys{} easy to use with various tools (e.g., the PISA engine~\cite{MSMS2019} can index using the document export). However, deeper integration can provide further functionality, as we demonstrate in this section with four tools. Note that \sys does not depend on any of these tools; instead they use \sys.


\textbf{Capreolus}~\cite{yates2020capreolus} is a toolkit for training and evaluating neural learning-to-rank models through Python and command line interfaces. In terms of data, it includes components for ``collections`` (sets of documents) and ``benchmarks'' (sets of queries and qrels). Though it has some built-in datasets, it also supports all datasets available from \sys in its pipelines:
\begin{minted}[fontsize=\small]{python}
import capreolus as cap
collection, benchmark = cap.get_irds("pmc/v2/trec-cds-2016",
                fields=["abstract"], query_type="summary")
index = cap.AnseriniIndex({"stemmer": None}, collection)
index.create_index()
benchmark.qrels
benchmark.queries
\end{minted}

\textbf{PyTerrier}~\cite{pyterrier2020ictir} is a Python interface to the Terrier search engine~\cite{ounis2005terrier} that enables the creation of flexible retrieval pipelines. It has a native dataset API, but it now also automatically adds all datasets from \sys, expanding the number of available datasets. They can be accessed via the dataset ID with an \texttt{irds:} prefix, and then used seamlessly with the rest of PyTerrier:
\begin{minted}[fontsize=\small]{python}
import pyterrier as pt
pt.init()
ds = pt.get_dataset('irds:cord19/trec-covid')
indexer = pt.index.IterDictIndexer('./cord19')
indexer.index(ds.get_corpus_iter(), fields=('abstract',))
topics = ds.get_topics(variant="description")
qrels = ds.get_qrels()
\end{minted}

\textbf{OpenNIR}~\cite{macavaney:wsdm2020-onir} provides a command line neural reranking pipeline for several standard IR benchmarks.
OpenNIR supports \sys for its training, validation, and testing dataset components. Queries and qrels are trivially fed into the training and validation processes. Documents are automatically indexed with Anserini for first-stage retrieval, and document lookups are used to fetch the text when training and scoring. Here is an example testing on the TREC COVID dataset:

\begin{minted}[fontsize=\small]{bash}
$ scripts/pipeline.sh test_ds=irds test_ds.ds=cord19/trec-covid
\end{minted}

\textbf{Anserini}~\cite{Yang2017AnseriniET}, and its Python-wrapper counterpart Pyserini~\cite{Lin2021PyseriniAE} focus on reproducibility in IR. They provide a wrapper and suite of tools around a Lucene index. As such, operations on datasets in this tool are tightly coupled with the Lucene and Anserini packages.
Though it has support for a wide variety of query and relevance assessments (distributed with the package), the support for document content is sparse, since only a few collections have automatically-downloadable indices. The remainder rely on manual instructions.
Queries and qrels from \sys can be used with Anserini by using the export CLI (as TSV or TREC format). The CLI can also efficiently output documents in a format it can index in parallel:
\begin{minted}[fontsize=\small]{bash}
$ ir_datasets doc_fifos medline/2017
# To index with Anserini, run:
# IndexCollection -collection JsonCollection -input 
#   /tmp/tmp6sope5gr -threads 23 -index <your_index_path>
\end{minted}

\textbf{DiffIR}~\cite{jose:sigir2021-diffir} is a tool that enables the visualization and qualitative comparison of search results. Using \sys{}, it shows the textual content of the top results for queries and highlights model-specific impactful text spans.

\section{Community Contributions}

We welcome (and encourage) community contributions. Extending \sys as a separate package is straightforward,\footnote{example: \url{https://github.com/seanmacavaney/dummy-irds-ext}} and we also welcome pull requests to the main package. 

To maintain quality in \sys, we require considerations of ease-of-use, efficiency, data integrity, and documentation. We request that issues are opened before implementation to ensure proper consideration of these aspects. \sys provides tools for handling typical data formats (e.g., TREC, TSV, CSV), making the process relatively straightforward. Atypical formats likely require special processing. There are plenty of examples to help guide the contributor.

\section{Future Applications}

We envision \sys enabling a variety of useful applications.

\textbf{Training/evaluation in private settings.} This tool could facilitate experiments and tasks that involve keeping data private. This is a realistic setting in several circumstances. For instance, a shared task involving searching through clinical notes would likely face challenges distributing this collection due to patient privacy concerns. Or a company may want to offer a shared task using a proprietary document collection or query log. In both these cases, a version of \sys could be built that provides this data that is only available in a secure environment (e.g., one where networking is disabled). Participants could feel confident that their code is processing the data correctly, given that it supports the \sys{} API; their code can switch to this dataset simply by using the dataset ID of the private dataset.

\textbf{Dataset exploration GUI.} Performing ad-hoc data analysis using \sys is an improvement over prior approaches. The user experience could be further improved through a graphical user interface that facilitate common dataset exploration tasks. For instance, this tool could graphically present the list of queries and link to the text of judged documents. Though this functionality is easy through the Python and command line interfaces, a graphical interface would further reduce friction and ease exploration.

\section{Conclusion}

We presented \sys, a tool that provides access to a variety of datasets and benchmarks for search engines. The tool automatically downloads and verifies content when possible, to aid in reproducibility. Through Python and command-line interfaces, users can iterate over documents, queries, and relevance judgments, and perform lookups of documents by ID. The utility of these functionalities are demonstrated through integration with several tools for performing IR experiments. The \sys catalog can help users discover datasets and acts as a hub of information with links and citations to relevant literature. We hope that \sys reduces researcher burden, helps reduce redundant copies of datasets across toolkits, and enables the creation of new tools.



\begin{acks}
We thank those who contributed to issues or discussions about \sys.
This was funded in part by the ARCS Foundation.
\end{acks}

\bibliographystyle{ACM-Reference-Format}
\bibliography{biblio}


\begin{thebibliography}{94}


\ifx \showCODEN    \undefined \def \showCODEN     #1{\unskip}     \fi
\ifx \showDOI      \undefined \def \showDOI       #1{#1}\fi
\ifx \showISBNx    \undefined \def \showISBNx     #1{\unskip}     \fi
\ifx \showISBNxiii \undefined \def \showISBNxiii  #1{\unskip}     \fi
\ifx \showISSN     \undefined \def \showISSN      #1{\unskip}     \fi
\ifx \showLCCN     \undefined \def \showLCCN      #1{\unskip}     \fi
\ifx \shownote     \undefined \def \shownote      #1{#1}          \fi
\ifx \showarticletitle \undefined \def \showarticletitle #1{#1}   \fi
\ifx \showURL      \undefined \def \showURL       {\relax}        \fi
\providecommand\bibfield[2]{#2}
\providecommand\bibinfo[2]{#2}
\providecommand\natexlab[1]{#1}
\providecommand\showeprint[2][]{arXiv:#2}

\bibitem[\protect\citeauthoryear{??}{cra}{[n.d.]}]%
        {cranfield}
 \bibinfo{year}{[n.d.]}\natexlab{}.
\newblock \bibinfo{booktitle}{\emph{Cranfield collection}}.
\newblock
\urldef\tempurl%
\url{http://ir.dcs.gla.ac.uk/resources/test_collections/cran/}
\showURL{%
\tempurl}


\bibitem[\protect\citeauthoryear{??}{vas}{[n.d.]}]%
        {vaswani}
 \bibinfo{year}{[n.d.]}\natexlab{}.
\newblock \bibinfo{booktitle}{\emph{NPL collection}}.
\newblock
\urldef\tempurl%
\url{http://ir.dcs.gla.ac.uk/resources/test_collections/npl/}
\showURL{%
\tempurl}


\bibitem[\protect\citeauthoryear{??}{TFD}{[n.d.]}]%
        {TFDS}
 \bibinfo{year}{[n.d.]}\natexlab{}.
\newblock \bibinfo{booktitle}{\emph{{TensorFlow Datasets}, A collection of
  ready-to-use datasets}}.
\newblock
\urldef\tempurl%
\url{https://www.tensorflow.org/datasets}
\showURL{%
\tempurl}


\bibitem[\protect\citeauthoryear{Abualsaud, Lioma, Maistro, Smucker, and
  Zuccon}{Abualsaud et~al\mbox{.}}{2019}]%
        {trecdecision2019}
\bibfield{author}{\bibinfo{person}{Mustafa Abualsaud},
  \bibinfo{person}{Christina Lioma}, \bibinfo{person}{Maria Maistro},
  \bibinfo{person}{Mark~D. Smucker}, {and} \bibinfo{person}{Guido Zuccon}.}
  \bibinfo{year}{2019}\natexlab{}.
\newblock \showarticletitle{Overview of the TREC 2019 Decision Track}. In
  \bibinfo{booktitle}{\emph{TREC}}.
\newblock


\bibitem[\protect\citeauthoryear{Allan, Aslam, Carterette, Pavlu, and
  Kanoulas}{Allan et~al\mbox{.}}{2008}]%
        {mq08}
\bibfield{author}{\bibinfo{person}{James Allan}, \bibinfo{person}{Javed~A.
  Aslam}, \bibinfo{person}{Ben Carterette}, \bibinfo{person}{Virgil Pavlu},
  {and} \bibinfo{person}{Evangelos Kanoulas}.} \bibinfo{year}{2008}\natexlab{}.
\newblock \showarticletitle{Million Query Track 2008 Overview}. In
  \bibinfo{booktitle}{\emph{TREC}}.
\newblock


\bibitem[\protect\citeauthoryear{Allan, Carterette, Aslam, Pavlu, Dachev, and
  Kanoulas}{Allan et~al\mbox{.}}{2007}]%
        {mq07}
\bibfield{author}{\bibinfo{person}{James Allan}, \bibinfo{person}{Ben
  Carterette}, \bibinfo{person}{Javed~A. Aslam}, \bibinfo{person}{Virgil
  Pavlu}, \bibinfo{person}{Blagovest Dachev}, {and} \bibinfo{person}{Evangelos
  Kanoulas}.} \bibinfo{year}{2007}\natexlab{}.
\newblock \showarticletitle{Million Query Track 2007 Overview}. In
  \bibinfo{booktitle}{\emph{TREC}}.
\newblock


\bibitem[\protect\citeauthoryear{Allan, Harman, Kanoulas, Li, Gysel, and
  Vorhees}{Allan et~al\mbox{.}}{2017}]%
        {allan2017treccore}
\bibfield{author}{\bibinfo{person}{James Allan}, \bibinfo{person}{Donna
  Harman}, \bibinfo{person}{Evangelos Kanoulas}, \bibinfo{person}{Dan Li},
  \bibinfo{person}{Christophe~Van Gysel}, {and} \bibinfo{person}{Ellen
  Vorhees}.} \bibinfo{year}{2017}\natexlab{}.
\newblock \showarticletitle{TREC 2017 Common Core Track Overview}. In
  \bibinfo{booktitle}{\emph{TREC}}.
\newblock


\bibitem[\protect\citeauthoryear{Bondarenko, Fr{\"o}be, Beloucif, Gienapp,
  Ajjour, Panchenko, Biemann, Stein, Wachsmuth, Potthast, and Hagen}{Bondarenko
  et~al\mbox{.}}{2020}]%
        {to}
\bibfield{author}{\bibinfo{person}{Alexander Bondarenko}, \bibinfo{person}{Maik
  Fr{\"o}be}, \bibinfo{person}{Meriem Beloucif}, \bibinfo{person}{Lukas
  Gienapp}, \bibinfo{person}{Yamen Ajjour}, \bibinfo{person}{Alexander
  Panchenko}, \bibinfo{person}{Christian Biemann}, \bibinfo{person}{Benno
  Stein}, \bibinfo{person}{Henning Wachsmuth}, \bibinfo{person}{Martin
  Potthast}, {and} \bibinfo{person}{Matthias Hagen}.}
  \bibinfo{year}{2020}\natexlab{}.
\newblock \showarticletitle{Overview of Touch{\'e} 2020: Argument Retrieval}.
  In \bibinfo{booktitle}{\emph{CLEF}}.
\newblock


\bibitem[\protect\citeauthoryear{Boteva, Gholipour, Sokolov, and
  Riezler}{Boteva et~al\mbox{.}}{2016}]%
        {boteva16full}
\bibfield{author}{\bibinfo{person}{Vera Boteva}, \bibinfo{person}{Demian
  Gholipour}, \bibinfo{person}{Artem Sokolov}, {and} \bibinfo{person}{Stefan
  Riezler}.} \bibinfo{year}{2016}\natexlab{}.
\newblock \showarticletitle{A Full-Text Learning to Rank Dataset for Medical
  Information Retrieval}. In \bibinfo{booktitle}{\emph{Proceedings of the
  European Conference on Information Retrieval ({ECIR})}} (Padova, Italy).
  \bibinfo{publisher}{Springer}.
\newblock


\bibitem[\protect\citeauthoryear{B\"uttcher, Clarke, and Soboroff}{B\"uttcher
  et~al\mbox{.}}{2006}]%
        {tb06}
\bibfield{author}{\bibinfo{person}{Stefan B\"uttcher}, \bibinfo{person}{Charles
  L.~A. Clarke}, {and} \bibinfo{person}{Ian Soboroff}.}
  \bibinfo{year}{2006}\natexlab{}.
\newblock \showarticletitle{The TREC 2006 Terabyte Track}. In
  \bibinfo{booktitle}{\emph{TREC}}.
\newblock


\bibitem[\protect\citeauthoryear{Carterette, Pavlu, Fang, and
  Kanoulas}{Carterette et~al\mbox{.}}{2009}]%
        {mq09}
\bibfield{author}{\bibinfo{person}{Ben Carterette}, \bibinfo{person}{Virgil
  Pavlu}, \bibinfo{person}{Hui Fang}, {and} \bibinfo{person}{Evangelos
  Kanoulas}.} \bibinfo{year}{2009}\natexlab{}.
\newblock \showarticletitle{Million Query Track 2009 Overview}. In
  \bibinfo{booktitle}{\emph{TREC}}.
\newblock


\bibitem[\protect\citeauthoryear{Clark, Scholer, and Soboroff}{Clark
  et~al\mbox{.}}{2005}]%
        {tb05}
\bibfield{author}{\bibinfo{person}{Charles L.~A. Clark}, \bibinfo{person}{Falk
  Scholer}, {and} \bibinfo{person}{Ian Soboroff}.}
  \bibinfo{year}{2005}\natexlab{}.
\newblock \showarticletitle{The TREC 2005 Terabyte Track}. In
  \bibinfo{booktitle}{\emph{TREC}}.
\newblock


\bibitem[\protect\citeauthoryear{Clarke, Craswell, and Soboroff}{Clarke
  et~al\mbox{.}}{2004}]%
        {tb04}
\bibfield{author}{\bibinfo{person}{Charles Clarke}, \bibinfo{person}{Nick
  Craswell}, {and} \bibinfo{person}{Ian Soboroff}.}
  \bibinfo{year}{2004}\natexlab{}.
\newblock \showarticletitle{Overview of the TREC 2004 Terabyte Track}. In
  \bibinfo{booktitle}{\emph{TREC}}.
\newblock


\bibitem[\protect\citeauthoryear{Clarke, Craswell, and Soboroff}{Clarke
  et~al\mbox{.}}{2009}]%
        {w09}
\bibfield{author}{\bibinfo{person}{Charles L.~A. Clarke}, \bibinfo{person}{Nick
  Craswell}, {and} \bibinfo{person}{Ian Soboroff}.}
  \bibinfo{year}{2009}\natexlab{}.
\newblock \showarticletitle{Overview of the TREC 2009 Web Track}. In
  \bibinfo{booktitle}{\emph{TREC}}.
\newblock


\bibitem[\protect\citeauthoryear{Clarke, Craswell, Soboroff, and
  Cormack}{Clarke et~al\mbox{.}}{2010}]%
        {w10}
\bibfield{author}{\bibinfo{person}{Charles L.~A. Clarke}, \bibinfo{person}{Nick
  Craswell}, \bibinfo{person}{Ian Soboroff}, {and} \bibinfo{person}{Gordon~V.
  Cormack}.} \bibinfo{year}{2010}\natexlab{}.
\newblock \showarticletitle{Overview of the TREC 2010 Web Track}. In
  \bibinfo{booktitle}{\emph{TREC}}.
\newblock


\bibitem[\protect\citeauthoryear{Clarke, Craswell, Soboroff, and
  Voorhees}{Clarke et~al\mbox{.}}{2011}]%
        {w11}
\bibfield{author}{\bibinfo{person}{Charles L.~A. Clarke}, \bibinfo{person}{Nick
  Craswell}, \bibinfo{person}{Ian Soboroff}, {and} \bibinfo{person}{Ellen~M.
  Voorhees}.} \bibinfo{year}{2011}\natexlab{}.
\newblock \showarticletitle{Overview of the TREC 2011 Web Track}. In
  \bibinfo{booktitle}{\emph{TREC}}.
\newblock


\bibitem[\protect\citeauthoryear{Clarke, Craswell, and Voorhees}{Clarke
  et~al\mbox{.}}{2012}]%
        {w12}
\bibfield{author}{\bibinfo{person}{Charles L.~A. Clarke}, \bibinfo{person}{Nick
  Craswell}, {and} \bibinfo{person}{Ellen~M. Voorhees}.}
  \bibinfo{year}{2012}\natexlab{}.
\newblock \showarticletitle{Overview of the TREC 2012 Web Track}. In
  \bibinfo{booktitle}{\emph{TREC}}.
\newblock


\bibitem[\protect\citeauthoryear{Cohan, Feldman, Beltagy, Downey, and
  Weld}{Cohan et~al\mbox{.}}{2020}]%
        {sd}
\bibfield{author}{\bibinfo{person}{Arman Cohan}, \bibinfo{person}{Sergey
  Feldman}, \bibinfo{person}{Iz Beltagy}, \bibinfo{person}{Doug Downey}, {and}
  \bibinfo{person}{Daniel Weld}.} \bibinfo{year}{2020}\natexlab{}.
\newblock \showarticletitle{{SPECTER}: Document-level Representation Learning
  using Citation-informed Transformers}. In
  \bibinfo{booktitle}{\emph{Proceedings of the 58th Annual Meeting of the
  Association for Computational Linguistics}}. \bibinfo{publisher}{Association
  for Computational Linguistics}, \bibinfo{address}{Online},
  \bibinfo{pages}{2270--2282}.
\newblock
\urldef\tempurl%
\url{https://doi.org/10.18653/v1/2020.acl-main.207}
\showDOI{\tempurl}


\bibitem[\protect\citeauthoryear{Collins-Thompson, Bennett, Diaz, Clarke, and
  Voorhees}{Collins-Thompson et~al\mbox{.}}{2013}]%
        {w13}
\bibfield{author}{\bibinfo{person}{Kevyn Collins-Thompson},
  \bibinfo{person}{Paul Bennett}, \bibinfo{person}{Fernando Diaz},
  \bibinfo{person}{Charles L.~A. Clarke}, {and} \bibinfo{person}{Ellen~M.
  Voorhees}.} \bibinfo{year}{2013}\natexlab{}.
\newblock \showarticletitle{TREC 2013 Web Track Overview}. In
  \bibinfo{booktitle}{\emph{TREC}}.
\newblock


\bibitem[\protect\citeauthoryear{Collins-Thompson, Macdonald, Bennett, Diaz,
  and Voorhees}{Collins-Thompson et~al\mbox{.}}{2014}]%
        {w14}
\bibfield{author}{\bibinfo{person}{Kevyn Collins-Thompson},
  \bibinfo{person}{Craig Macdonald}, \bibinfo{person}{Paul Bennett},
  \bibinfo{person}{Fernando Diaz}, {and} \bibinfo{person}{Ellen~M. Voorhees}.}
  \bibinfo{year}{2014}\natexlab{}.
\newblock \showarticletitle{TREC 2014 Web Track Overview}. In
  \bibinfo{booktitle}{\emph{TREC}}.
\newblock


\bibitem[\protect\citeauthoryear{Craswell, Campos, Mitra, Yilmaz, and
  Billerbeck}{Craswell et~al\mbox{.}}{2020a}]%
        {craswell2020orcas}
\bibfield{author}{\bibinfo{person}{Nick Craswell}, \bibinfo{person}{Daniel
  Campos}, \bibinfo{person}{Bhaskar Mitra}, \bibinfo{person}{Emine Yilmaz},
  {and} \bibinfo{person}{Bodo Billerbeck}.} \bibinfo{year}{2020}\natexlab{a}.
\newblock \showarticletitle{ORCAS: 18 Million Clicked Query-Document Pairs for
  Analyzing Search}.
\newblock \bibinfo{journal}{\emph{arXiv preprint arXiv:2006.05324}}
  (\bibinfo{year}{2020}).
\newblock


\bibitem[\protect\citeauthoryear{Craswell and Hawking}{Craswell and
  Hawking}{2002}]%
        {w02}
\bibfield{author}{\bibinfo{person}{Nick Craswell} {and} \bibinfo{person}{David
  Hawking}.} \bibinfo{year}{2002}\natexlab{}.
\newblock \showarticletitle{Overview of the TREC-2002 Web Track}. In
  \bibinfo{booktitle}{\emph{TREC}}.
\newblock


\bibitem[\protect\citeauthoryear{Craswell and Hawking}{Craswell and
  Hawking}{2004}]%
        {w04}
\bibfield{author}{\bibinfo{person}{Nick Craswell} {and} \bibinfo{person}{David
  Hawking}.} \bibinfo{year}{2004}\natexlab{}.
\newblock \showarticletitle{Overview of the TREC-2004 Web Track}. In
  \bibinfo{booktitle}{\emph{TREC}}.
\newblock


\bibitem[\protect\citeauthoryear{Craswell, Hawking, Wilkinson, and Wu}{Craswell
  et~al\mbox{.}}{2003}]%
        {w03}
\bibfield{author}{\bibinfo{person}{Nick Craswell}, \bibinfo{person}{David
  Hawking}, \bibinfo{person}{Ross Wilkinson}, {and} \bibinfo{person}{Mingfang
  Wu}.} \bibinfo{year}{2003}\natexlab{}.
\newblock \showarticletitle{Overview of the TREC 2003 Web Track}. In
  \bibinfo{booktitle}{\emph{TREC}}.
\newblock


\bibitem[\protect\citeauthoryear{Craswell, Mitra, Yilmaz, and Campos}{Craswell
  et~al\mbox{.}}{2020b}]%
        {dl20}
\bibfield{author}{\bibinfo{person}{Nick Craswell}, \bibinfo{person}{Bhaskar
  Mitra}, \bibinfo{person}{Emine Yilmaz}, {and} \bibinfo{person}{Daniel
  Campos}.} \bibinfo{year}{2020}\natexlab{b}.
\newblock \showarticletitle{Overview of the TREC 2020 deep learning track}. In
  \bibinfo{booktitle}{\emph{TREC}}.
\newblock


\bibitem[\protect\citeauthoryear{Craswell, Mitra, Yilmaz, Campos, and
  Voorhees}{Craswell et~al\mbox{.}}{2019}]%
        {dl19}
\bibfield{author}{\bibinfo{person}{Nick Craswell}, \bibinfo{person}{Bhaskar
  Mitra}, \bibinfo{person}{Emine Yilmaz}, \bibinfo{person}{Daniel Campos},
  {and} \bibinfo{person}{Ellen Voorhees}.} \bibinfo{year}{2019}\natexlab{}.
\newblock \showarticletitle{Overview of the TREC 2019 deep learning track}. In
  \bibinfo{booktitle}{\emph{TREC 2019}}.
\newblock


\bibitem[\protect\citeauthoryear{Devlin, Chang, Lee, and Toutanova}{Devlin
  et~al\mbox{.}}{2019}]%
        {DevlinCLT19}
\bibfield{author}{\bibinfo{person}{Jacob Devlin}, \bibinfo{person}{Ming{-}Wei
  Chang}, \bibinfo{person}{Kenton Lee}, {and} \bibinfo{person}{Kristina
  Toutanova}.} \bibinfo{year}{2019}\natexlab{}.
\newblock \showarticletitle{{BERT:} Pre-training of Deep Bidirectional
  Transformers for Language Understanding}. In
  \bibinfo{booktitle}{\emph{{NAACL-HLT}}}.
\newblock


\bibitem[\protect\citeauthoryear{Dietz and Gamari}{Dietz and Gamari}{2017}]%
        {Dietz2017}
\bibfield{author}{\bibinfo{person}{Laura Dietz} {and} \bibinfo{person}{Ben
  Gamari}.} \bibinfo{year}{2017}\natexlab{}.
\newblock \showarticletitle{{TREC CAR}: A Data Set for Complex Answer
  Retrieval}.
\newblock  (\bibinfo{year}{2017}).
\newblock
\urldef\tempurl%
\url{http://trec-car.cs.unh.edu}
\showURL{%
\tempurl}
\newblock
\shownote{Version 1.5.}


\bibitem[\protect\citeauthoryear{Dietz, Verma, Radlinski, and Craswell}{Dietz
  et~al\mbox{.}}{2017}]%
        {dietz2017trec}
\bibfield{author}{\bibinfo{person}{Laura Dietz}, \bibinfo{person}{Manisha
  Verma}, \bibinfo{person}{Filip Radlinski}, {and} \bibinfo{person}{Nick
  Craswell}.} \bibinfo{year}{2017}\natexlab{}.
\newblock \showarticletitle{TREC Complex Answer Retrieval Overview.}. In
  \bibinfo{booktitle}{\emph{TREC}}.
\newblock


\bibitem[\protect\citeauthoryear{Diggelmann, Boyd-Graber, Bulian, Ciaramita,
  and Leippold}{Diggelmann et~al\mbox{.}}{2020}]%
        {cf}
\bibfield{author}{\bibinfo{person}{T. Diggelmann}, \bibinfo{person}{Jordan~L.
  Boyd-Graber}, \bibinfo{person}{Jannis Bulian}, \bibinfo{person}{Massimiliano
  Ciaramita}, {and} \bibinfo{person}{Markus Leippold}.}
  \bibinfo{year}{2020}\natexlab{}.
\newblock \showarticletitle{CLIMATE-FEVER: A Dataset for Verification of
  Real-World Climate Claims}.
\newblock \bibinfo{journal}{\emph{ArXiv}}  \bibinfo{volume}{abs/2012.00614}
  (\bibinfo{year}{2020}).
\newblock


\bibitem[\protect\citeauthoryear{Frej, Schwab, and Chevallet}{Frej
  et~al\mbox{.}}{2020a}]%
        {mlwikir}
\bibfield{author}{\bibinfo{person}{Jibril Frej}, \bibinfo{person}{Didier
  Schwab}, {and} \bibinfo{person}{Jean-Pierre Chevallet}.}
  \bibinfo{year}{2020}\natexlab{a}.
\newblock \showarticletitle{MLWIKIR: A Python Toolkit for Building Large-scale
  Wikipedia-based Information Retrieval Datasets in Chinese, English, French,
  Italian, Japanese, Spanish and More}. In \bibinfo{booktitle}{\emph{CIRCLE}}.
\newblock


\bibitem[\protect\citeauthoryear{Frej, Schwab, and Chevallet}{Frej
  et~al\mbox{.}}{2020b}]%
        {wikir}
\bibfield{author}{\bibinfo{person}{Jibril Frej}, \bibinfo{person}{Didier
  Schwab}, {and} \bibinfo{person}{Jean-Pierre Chevallet}.}
  \bibinfo{year}{2020}\natexlab{b}.
\newblock \showarticletitle{WIKIR: A Python toolkit for building a large-scale
  Wikipedia-based English Information Retrieval Dataset}. In
  \bibinfo{booktitle}{\emph{LREC}}.
\newblock


\bibitem[\protect\citeauthoryear{Gey and Oard}{Gey and Oard}{2001}]%
        {arabic01}
\bibfield{author}{\bibinfo{person}{Fredric Gey} {and} \bibinfo{person}{Douglas
  Oard}.} \bibinfo{year}{2001}\natexlab{}.
\newblock \showarticletitle{The TREC-2001 Cross-Language Information Retrieval
  Track: Searching Arabic using English, French or Arabic Queries}. In
  \bibinfo{booktitle}{\emph{TREC}}.
\newblock


\bibitem[\protect\citeauthoryear{Gey and Oard}{Gey and Oard}{2002}]%
        {arabic02}
\bibfield{author}{\bibinfo{person}{Fredric Gey} {and} \bibinfo{person}{Douglas
  Oard}.} \bibinfo{year}{2002}\natexlab{}.
\newblock \showarticletitle{The TREC-2002 Arabic/English CLIR Track}. In
  \bibinfo{booktitle}{\emph{TREC}}.
\newblock


\bibitem[\protect\citeauthoryear{Graff and Walker}{Graff and Walker}{2001}]%
        {LDC2001T55}
\bibfield{author}{\bibinfo{person}{David Graff} {and} \bibinfo{person}{Kevin
  Walker}.} \bibinfo{year}{2001}\natexlab{}.
\newblock \bibinfo{title}{Arabic Newswire Part 1 LDC2001T55}.
\newblock
\newblock
\urldef\tempurl%
\url{https://catalog.ldc.upenn.edu/LDC2001T55}
\showURL{%
\tempurl}


\bibitem[\protect\citeauthoryear{Harman}{Harman}{1994}]%
        {es3}
\bibfield{author}{\bibinfo{person}{Donna Harman}.}
  \bibinfo{year}{1994}\natexlab{}.
\newblock \showarticletitle{Overview of the Third Text REtrieval Conference
  (TREC-3)}. In \bibinfo{booktitle}{\emph{TREC}}.
\newblock


\bibitem[\protect\citeauthoryear{Harman}{Harman}{1995}]%
        {es4}
\bibfield{author}{\bibinfo{person}{Donna Harman}.}
  \bibinfo{year}{1995}\natexlab{}.
\newblock \showarticletitle{Overview of the Fourth Text REtrieval Conference
  (TREC-4)}. In \bibinfo{booktitle}{\emph{TREC}}.
\newblock


\bibitem[\protect\citeauthoryear{Hashemi, Aliannejadi, Zamani, and
  Croft}{Hashemi et~al\mbox{.}}{2020}]%
        {Hashemi2020Antique}
\bibfield{author}{\bibinfo{person}{Helia Hashemi}, \bibinfo{person}{Mohammad
  Aliannejadi}, \bibinfo{person}{Hamed Zamani}, {and} \bibinfo{person}{Bruce
  Croft}.} \bibinfo{year}{2020}\natexlab{}.
\newblock \showarticletitle{ANTIQUE: A Non-Factoid Question Answering
  Benchmark}. In \bibinfo{booktitle}{\emph{ECIR}}.
\newblock


\bibitem[\protect\citeauthoryear{Hasibi, Nikolaev, Xiong, Balog, Bratsberg,
  Kotov, and Callan}{Hasibi et~al\mbox{.}}{2017}]%
        {dbp}
\bibfield{author}{\bibinfo{person}{Faegheh Hasibi}, \bibinfo{person}{Fedor
  Nikolaev}, \bibinfo{person}{Chenyan Xiong}, \bibinfo{person}{K. Balog},
  \bibinfo{person}{S.~E. Bratsberg}, \bibinfo{person}{Alexander Kotov}, {and}
  \bibinfo{person}{J. Callan}.} \bibinfo{year}{2017}\natexlab{}.
\newblock \showarticletitle{DBpedia-Entity v2: A Test Collection for Entity
  Search}.
\newblock \bibinfo{journal}{\emph{Proceedings of the 40th International ACM
  SIGIR Conference on Research and Development in Information Retrieval}}
  (\bibinfo{year}{2017}).
\newblock


\bibitem[\protect\citeauthoryear{Hersh, Cohen, Ruslen, and Roberts}{Hersh
  et~al\mbox{.}}{2007a}]%
        {gen07}
\bibfield{author}{\bibinfo{person}{William Hersh}, \bibinfo{person}{Aaron
  Cohen}, \bibinfo{person}{Lynn Ruslen}, {and} \bibinfo{person}{Phoebe
  Roberts}.} \bibinfo{year}{2007}\natexlab{a}.
\newblock \showarticletitle{TREC 2007 Genomics Track Overview}. In
  \bibinfo{booktitle}{\emph{TREC}}.
\newblock


\bibitem[\protect\citeauthoryear{Hersh, Cohen, Yang, Bhupatiraju, Roberts, and
  Hearst}{Hersh et~al\mbox{.}}{2007b}]%
        {gen05}
\bibfield{author}{\bibinfo{person}{William Hersh}, \bibinfo{person}{Aaron
  Cohen}, \bibinfo{person}{Jianji Yang}, \bibinfo{person}{Ravi~Teja
  Bhupatiraju}, \bibinfo{person}{Phoebe Roberts}, {and} \bibinfo{person}{Marti
  Hearst}.} \bibinfo{year}{2007}\natexlab{b}.
\newblock \showarticletitle{TREC 2005 Genomics Track Overview}. In
  \bibinfo{booktitle}{\emph{TREC}}.
\newblock


\bibitem[\protect\citeauthoryear{Hersh, Cohen, Roberts, and Rekapalli}{Hersh
  et~al\mbox{.}}{2006}]%
        {gen06}
\bibfield{author}{\bibinfo{person}{William Hersh}, \bibinfo{person}{Aaron~M.
  Cohen}, \bibinfo{person}{Phoebe Roberts}, {and} \bibinfo{person}{Hari~Krishna
  Rekapalli}.} \bibinfo{year}{2006}\natexlab{}.
\newblock \showarticletitle{TREC 2006 Genomics Track Overview}. In
  \bibinfo{booktitle}{\emph{TREC}}.
\newblock


\bibitem[\protect\citeauthoryear{Hersh, Bhuptiraju, Ross, Johnson, Cohen, and
  Kraemer}{Hersh et~al\mbox{.}}{2004}]%
        {gen04}
\bibfield{author}{\bibinfo{person}{William~R. Hersh},
  \bibinfo{person}{Ravi~Teja Bhuptiraju}, \bibinfo{person}{Laura Ross},
  \bibinfo{person}{Phoebe Johnson}, \bibinfo{person}{Aaron~M. Cohen}, {and}
  \bibinfo{person}{Dale~F. Kraemer}.} \bibinfo{year}{2004}\natexlab{}.
\newblock \showarticletitle{TREC 2004 Genomics Track Overview}. In
  \bibinfo{booktitle}{\emph{TREC}}.
\newblock


\bibitem[\protect\citeauthoryear{Hoogeveen, Verspoor, and Baldwin}{Hoogeveen
  et~al\mbox{.}}{2015}]%
        {cqa}
\bibfield{author}{\bibinfo{person}{D. Hoogeveen}, \bibinfo{person}{Karin~M.
  Verspoor}, {and} \bibinfo{person}{Timothy Baldwin}.}
  \bibinfo{year}{2015}\natexlab{}.
\newblock \showarticletitle{{CQADupStack}: A Benchmark Data Set for Community
  Question-Answering Research}.
\newblock \bibinfo{journal}{\emph{Proceedings of the 20th Australasian Document
  Computing Symposium}} (\bibinfo{year}{2015}).
\newblock


\bibitem[\protect\citeauthoryear{Husain, Wu, Gazit, Allamanis, and
  Brockschmidt}{Husain et~al\mbox{.}}{2019}]%
        {Husain2019CodeSearchNet}
\bibfield{author}{\bibinfo{person}{Hamel Husain}, \bibinfo{person}{Ho-Hsiang
  Wu}, \bibinfo{person}{Tiferet Gazit}, \bibinfo{person}{Miltiadis Allamanis},
  {and} \bibinfo{person}{Marc Brockschmidt}.} \bibinfo{year}{2019}\natexlab{}.
\newblock \showarticletitle{CodeSearchNet Challenge: Evaluating the State of
  Semantic Code Search}.
\newblock \bibinfo{journal}{\emph{ArXiv}} (\bibinfo{year}{2019}).
\newblock


\bibitem[\protect\citeauthoryear{Jose, Nguyen, MacAvaney, Dalton, and
  Yates}{Jose et~al\mbox{.}}{2021}]%
        {jose:sigir2021-diffir}
\bibfield{author}{\bibinfo{person}{Kevin~Martin Jose}, \bibinfo{person}{Thong
  Nguyen}, \bibinfo{person}{Sean MacAvaney}, \bibinfo{person}{Jeff Dalton},
  {and} \bibinfo{person}{Andrew Yates}.} \bibinfo{year}{2021}\natexlab{}.
\newblock \showarticletitle{DiffIR: Exploring Differences in Ranking Models'
  Behavior}. In \bibinfo{booktitle}{\emph{Proceedings of the 44th International
  ACM SIGIR Conference on Research and Development in Information Retrieval}}.
\newblock


\bibitem[\protect\citeauthoryear{Joshi, Choi, Weld, and Zettlemoyer}{Joshi
  et~al\mbox{.}}{2017}]%
        {Joshi2017TriviaQAAL}
\bibfield{author}{\bibinfo{person}{Mandar Joshi}, \bibinfo{person}{Eunsol
  Choi}, \bibinfo{person}{Daniel~S. Weld}, {and} \bibinfo{person}{Luke
  Zettlemoyer}.} \bibinfo{year}{2017}\natexlab{}.
\newblock \showarticletitle{TriviaQA: A Large Scale Distantly Supervised
  Challenge Dataset for Reading Comprehension}. In
  \bibinfo{booktitle}{\emph{ACL}}.
\newblock


\bibitem[\protect\citeauthoryear{Karpukhin, Oğuz, Min, Lewis, Wu, Edunov,
  Chen, and tau Yih}{Karpukhin et~al\mbox{.}}{2020}]%
        {dprnq}
\bibfield{author}{\bibinfo{person}{Vladimir Karpukhin}, \bibinfo{person}{Barlas
  Oğuz}, \bibinfo{person}{Sewon Min}, \bibinfo{person}{Patrick Lewis},
  \bibinfo{person}{Ledell Wu}, \bibinfo{person}{Sergey Edunov},
  \bibinfo{person}{Danqi Chen}, {and} \bibinfo{person}{Wen tau Yih}.}
  \bibinfo{year}{2020}\natexlab{}.
\newblock \bibinfo{title}{Dense Passage Retrieval for Open-Domain Question
  Answering}.
\newblock
\newblock
\showeprint[arxiv]{2004.04906}~[cs.CL]


\bibitem[\protect\citeauthoryear{Khattab and Zaharia}{Khattab and
  Zaharia}{2020}]%
        {Khattab2020ColBERTEA}
\bibfield{author}{\bibinfo{person}{Omar Khattab} {and} \bibinfo{person}{Matei
  Zaharia}.} \bibinfo{year}{2020}\natexlab{}.
\newblock \showarticletitle{ColBERT: Efficient and Effective Passage Search via
  Contextualized Late Interaction over BERT}.
\newblock \bibinfo{journal}{\emph{SIGIR}} (\bibinfo{year}{2020}).
\newblock


\bibitem[\protect\citeauthoryear{Kwiatkowski, Palomaki, Redfield, Collins,
  Parikh, Alberti, Epstein, Polosukhin, Kelcey, Devlin, Lee, Toutanova, Jones,
  Chang, Dai, Uszkoreit, Le, and Petrov}{Kwiatkowski et~al\mbox{.}}{2019}]%
        {nq}
\bibfield{author}{\bibinfo{person}{Tom Kwiatkowski},
  \bibinfo{person}{Jennimaria Palomaki}, \bibinfo{person}{Olivia Redfield},
  \bibinfo{person}{Michael Collins}, \bibinfo{person}{Ankur Parikh},
  \bibinfo{person}{Chris Alberti}, \bibinfo{person}{Danielle Epstein},
  \bibinfo{person}{Illia Polosukhin}, \bibinfo{person}{Matthew Kelcey},
  \bibinfo{person}{Jacob Devlin}, \bibinfo{person}{Kenton Lee},
  \bibinfo{person}{Kristina~N. Toutanova}, \bibinfo{person}{Llion Jones},
  \bibinfo{person}{Ming-Wei Chang}, \bibinfo{person}{Andrew Dai},
  \bibinfo{person}{Jakob Uszkoreit}, \bibinfo{person}{Quoc Le}, {and}
  \bibinfo{person}{Slav Petrov}.} \bibinfo{year}{2019}\natexlab{}.
\newblock \showarticletitle{Natural Questions: a Benchmark for Question
  Answering Research}.
\newblock \bibinfo{journal}{\emph{TACL}} (\bibinfo{year}{2019}).
\newblock


\bibitem[\protect\citeauthoryear{Lin and Efron}{Lin and Efron}{2013}]%
        {mb13}
\bibfield{author}{\bibinfo{person}{Jimmy Lin} {and} \bibinfo{person}{Miles
  Efron}.} \bibinfo{year}{2013}\natexlab{}.
\newblock \showarticletitle{Overview of the TREC-2013 Microblog Track}. In
  \bibinfo{booktitle}{\emph{TREC}}.
\newblock


\bibitem[\protect\citeauthoryear{Lin, Efron, Wang, and Sherman}{Lin
  et~al\mbox{.}}{2014}]%
        {mb14}
\bibfield{author}{\bibinfo{person}{Jimmy Lin}, \bibinfo{person}{Miles Efron},
  \bibinfo{person}{Yulu Wang}, {and} \bibinfo{person}{Garrick Sherman}.}
  \bibinfo{year}{2014}\natexlab{}.
\newblock \showarticletitle{Overview of the TREC-2014 Microblog Track}. In
  \bibinfo{booktitle}{\emph{TREC}}.
\newblock


\bibitem[\protect\citeauthoryear{Lin, Ma, Lin, Yang, Pradeep, and Nogueira}{Lin
  et~al\mbox{.}}{2021}]%
        {Lin2021PyseriniAE}
\bibfield{author}{\bibinfo{person}{Jimmy Lin}, \bibinfo{person}{Xueguang Ma},
  \bibinfo{person}{Sheng-Chieh Lin}, \bibinfo{person}{Jheng-Hong Yang},
  \bibinfo{person}{Ronak Pradeep}, {and} \bibinfo{person}{Rodrigo Nogueira}.}
  \bibinfo{year}{2021}\natexlab{}.
\newblock \showarticletitle{Pyserini: An Easy-to-Use Python Toolkit to Support
  Replicable IR Research with Sparse and Dense Representations}.
\newblock \bibinfo{journal}{\emph{ArXiv}}  \bibinfo{volume}{abs/2102.10073}
  (\bibinfo{year}{2021}).
\newblock


\bibitem[\protect\citeauthoryear{Luo, Sakai, Liu, Dou, Xiong, and Xu}{Luo
  et~al\mbox{.}}{2017}]%
        {ntcirwww1}
\bibfield{author}{\bibinfo{person}{Cheng Luo}, \bibinfo{person}{Tetsuya Sakai},
  \bibinfo{person}{Yiqun Liu}, \bibinfo{person}{Zhicheng Dou},
  \bibinfo{person}{Chenyan Xiong}, {and} \bibinfo{person}{Jingfang Xu}.}
  \bibinfo{year}{2017}\natexlab{}.
\newblock \showarticletitle{Overview of the NTCIR-13 We Want Web Task}. In
  \bibinfo{booktitle}{\emph{NTCIR}}.
\newblock


\bibitem[\protect\citeauthoryear{MacAvaney}{MacAvaney}{2020}]%
        {macavaney:wsdm2020-onir}
\bibfield{author}{\bibinfo{person}{Sean MacAvaney}.}
  \bibinfo{year}{2020}\natexlab{}.
\newblock \showarticletitle{OpenNIR: A Complete Neural Ad-Hoc Ranking
  Pipeline}. In \bibinfo{booktitle}{\emph{Proceedings of the Thirteenth ACM
  International Conference on Web Search and Data Mining}}.
  \bibinfo{pages}{845--848}.
\newblock
\urldef\tempurl%
\url{https://doi.org/10.1145/3336191.3371864}
\showDOI{\tempurl}


\bibitem[\protect\citeauthoryear{MacAvaney, Nardini, Perego, Tonellotto,
  Goharian, and Frieder}{MacAvaney et~al\mbox{.}}{2020}]%
        {macavaney:sigir2020-eff}
\bibfield{author}{\bibinfo{person}{Sean MacAvaney},
  \bibinfo{person}{Franco~Maria Nardini}, \bibinfo{person}{Raffaele Perego},
  \bibinfo{person}{Nicola Tonellotto}, \bibinfo{person}{Nazli Goharian}, {and}
  \bibinfo{person}{Ophir Frieder}.} \bibinfo{year}{2020}\natexlab{}.
\newblock \showarticletitle{Efficient Document Re-Ranking for Transformers by
  Precomputing Term Representations}. In \bibinfo{booktitle}{\emph{Proceedings
  of the 43rd International ACM SIGIR Conference on Research and Development in
  Information Retrieval}}. \bibinfo{pages}{49--58}.
\newblock
\urldef\tempurl%
\url{https://doi.org/10.1145/3397271.3401093}
\showDOI{\tempurl}


\bibitem[\protect\citeauthoryear{MacAvaney, Yates, Hui, and Frieder}{MacAvaney
  et~al\mbox{.}}{2019}]%
        {macavaney:nyt}
\bibfield{author}{\bibinfo{person}{Sean MacAvaney}, \bibinfo{person}{Andrew
  Yates}, \bibinfo{person}{Kai Hui}, {and} \bibinfo{person}{Ophir Frieder}.}
  \bibinfo{year}{2019}\natexlab{}.
\newblock \showarticletitle{Content-Based Weak Supervision for Ad-Hoc
  Re-Ranking}. In \bibinfo{booktitle}{\emph{SIGIR}}.
\newblock


\bibitem[\protect\citeauthoryear{Macdonald and Tonellotto}{Macdonald and
  Tonellotto}{2020}]%
        {pyterrier2020ictir}
\bibfield{author}{\bibinfo{person}{Craig Macdonald} {and}
  \bibinfo{person}{Nicola Tonellotto}.} \bibinfo{year}{2020}\natexlab{}.
\newblock \showarticletitle{Declarative Experimentation inInformation Retrieval
  using PyTerrier}. In \bibinfo{booktitle}{\emph{Proceedings of ICTIR 2020}}.
\newblock


\bibitem[\protect\citeauthoryear{Mackie, Dalton, and Yates}{Mackie
  et~al\mbox{.}}{2021}]%
        {Mackie2021-howdeep}
\bibfield{author}{\bibinfo{person}{Iain Mackie}, \bibinfo{person}{Jeffrey
  Dalton}, {and} \bibinfo{person}{Andrew Yates}.}
  \bibinfo{year}{2021}\natexlab{}.
\newblock \showarticletitle{How Deep is your Learning: the DL-HARD Annotated
  Deep Learning Dataset}. In \bibinfo{booktitle}{\emph{Proceedings of the 44th
  International ACM SIGIR Conference on Research and Development in Information
  Retrieval}}.
\newblock


\bibitem[\protect\citeauthoryear{Maia, Handschuh, Freitas, Davis, McDermott,
  Zarrouk, and Balahur}{Maia et~al\mbox{.}}{2018}]%
        {fi}
\bibfield{author}{\bibinfo{person}{Macedo Maia}, \bibinfo{person}{S.
  Handschuh}, \bibinfo{person}{A. Freitas}, \bibinfo{person}{Brian Davis},
  \bibinfo{person}{R. McDermott}, \bibinfo{person}{M. Zarrouk}, {and}
  \bibinfo{person}{A. Balahur}.} \bibinfo{year}{2018}\natexlab{}.
\newblock \showarticletitle{WWW'18 Open Challenge: Financial Opinion Mining and
  Question Answering}.
\newblock \bibinfo{journal}{\emph{Companion Proceedings of the The Web
  Conference 2018}} (\bibinfo{year}{2018}).
\newblock


\bibitem[\protect\citeauthoryear{Mallia, Siedlaczek, Mackenzie, and
  Suel}{Mallia et~al\mbox{.}}{2019}]%
        {MSMS2019}
\bibfield{author}{\bibinfo{person}{Antonio Mallia}, \bibinfo{person}{Michal
  Siedlaczek}, \bibinfo{person}{Joel Mackenzie}, {and} \bibinfo{person}{Torsten
  Suel}.} \bibinfo{year}{2019}\natexlab{}.
\newblock \showarticletitle{{PISA:} Performant Indexes and Search for
  Academia}. In \bibinfo{booktitle}{\emph{Proceedings of the Open-Source {IR}
  Replicability Challenge co-located with 42nd International {ACM} {SIGIR}
  Conference on Research and Development in Information Retrieval, OSIRRC@SIGIR
  2019, Paris, France, July 25, 2019.}} \bibinfo{pages}{50--56}.
\newblock
\urldef\tempurl%
\url{http://ceur-ws.org/Vol-2409/docker08.pdf}
\showURL{%
\tempurl}


\bibitem[\protect\citeauthoryear{Mao, Sakai, Luo, Xiao, Liu, and Dou}{Mao
  et~al\mbox{.}}{2018}]%
        {ntcirwww2}
\bibfield{author}{\bibinfo{person}{Jiaxin Mao}, \bibinfo{person}{Tetsuya
  Sakai}, \bibinfo{person}{Cheng Luo}, \bibinfo{person}{Peng Xiao},
  \bibinfo{person}{Yiqun Liu}, {and} \bibinfo{person}{Zhicheng Dou}.}
  \bibinfo{year}{2018}\natexlab{}.
\newblock \showarticletitle{Overview of the NTCIR-14 We Want Web Task}. In
  \bibinfo{booktitle}{\emph{NTCIR}}.
\newblock


\bibitem[\protect\citeauthoryear{Ounis, Amati, Plachouras, He, Macdonald, and
  Johnson}{Ounis et~al\mbox{.}}{2005}]%
        {ounis2005terrier}
\bibfield{author}{\bibinfo{person}{Iadh Ounis}, \bibinfo{person}{Gianni Amati},
  \bibinfo{person}{Vassilis Plachouras}, \bibinfo{person}{Ben He},
  \bibinfo{person}{Craig Macdonald}, {and} \bibinfo{person}{Douglas Johnson}.}
  \bibinfo{year}{2005}\natexlab{}.
\newblock \showarticletitle{Terrier information retrieval platform}. In
  \bibinfo{booktitle}{\emph{ECIR}}. Springer, \bibinfo{pages}{517--519}.
\newblock


\bibitem[\protect\citeauthoryear{Palotti, Zuccon, Jimmy, Pecina, Lupu,
  Goeuriot, Kelly, and Hanbury}{Palotti et~al\mbox{.}}{2017}]%
        {Palotti2017CLEF}
\bibfield{author}{\bibinfo{person}{Joao Palotti}, \bibinfo{person}{Guido
  Zuccon}, \bibinfo{person}{Jimmy}, \bibinfo{person}{Pavel Pecina},
  \bibinfo{person}{Mihai Lupu}, \bibinfo{person}{Lorraine Goeuriot},
  \bibinfo{person}{Liadh Kelly}, {and} \bibinfo{person}{Allan Hanbury}.}
  \bibinfo{year}{2017}\natexlab{}.
\newblock \showarticletitle{CLEF 2017 Task Overview: The IR Task at the eHealth
  Evaluation Lab - Evaluating Retrieval Methods for Consumer Health Search}. In
  \bibinfo{booktitle}{\emph{CLEF}}.
\newblock


\bibitem[\protect\citeauthoryear{pandas~development team}{pandas~development
  team}{2020}]%
        {reback2020pandas}
\bibfield{author}{\bibinfo{person}{The pandas~development team}.}
  \bibinfo{year}{2020}\natexlab{}.
\newblock \bibinfo{booktitle}{\emph{pandas-dev/pandas: Pandas}}.
\newblock
\urldef\tempurl%
\url{https://doi.org/10.5281/zenodo.3509134}
\showDOI{\tempurl}


\bibitem[\protect\citeauthoryear{Payal~Bajaj}{Payal~Bajaj}{2016}]%
        {Bajaj2016MSMA}
\bibfield{author}{\bibinfo{person}{Nick Craswell Li Deng Jianfeng Gao Xiaodong
  Liu Rangan Majumder Andrew McNamara Bhaskar Mitra Tri Nguyen Mir Rosenberg
  Xia Song Alina Stoica Saurabh Tiwary Tong~Wang Payal~Bajaj, Daniel~Campos}.}
  \bibinfo{year}{2016}\natexlab{}.
\newblock \showarticletitle{MS MARCO: A Human Generated MAchine Reading
  COmprehension Dataset}. In \bibinfo{booktitle}{\emph{InCoCo@NIPS}}.
\newblock


\bibitem[\protect\citeauthoryear{Rekabsaz, Lesota, Schedl, Brassey, and
  Eickhoff}{Rekabsaz et~al\mbox{.}}{2021}]%
        {rekabsaz2021tripclick}
\bibfield{author}{\bibinfo{person}{Navid Rekabsaz}, \bibinfo{person}{Oleg
  Lesota}, \bibinfo{person}{Markus Schedl}, \bibinfo{person}{Jon Brassey},
  {and} \bibinfo{person}{Carsten Eickhoff}.} \bibinfo{year}{2021}\natexlab{}.
\newblock \showarticletitle{TripClick: The Log Files of a Large Health Web
  Search Engine}. In \bibinfo{booktitle}{\emph{SIGIR}}.
\newblock


\bibitem[\protect\citeauthoryear{Roberts, Demner-Fushman, Voorhees, Hersh,
  Bedrick, and Lazar}{Roberts et~al\mbox{.}}{2018}]%
        {pm18}
\bibfield{author}{\bibinfo{person}{Kirk Roberts}, \bibinfo{person}{Dina
  Demner-Fushman}, \bibinfo{person}{Ellen Voorhees},
  \bibinfo{person}{William~R. Hersh}, \bibinfo{person}{Steven Bedrick}, {and}
  \bibinfo{person}{Alexander~J. Lazar}.} \bibinfo{year}{2018}\natexlab{}.
\newblock \showarticletitle{Overview of the TREC 2018 Precision Medicine
  Track}. In \bibinfo{booktitle}{\emph{TREC}}.
\newblock


\bibitem[\protect\citeauthoryear{Roberts, Demner-Fushman, Voorhees, Hersh,
  Bedrick, Lazar, and Pant}{Roberts et~al\mbox{.}}{2017}]%
        {pm17}
\bibfield{author}{\bibinfo{person}{Kirk Roberts}, \bibinfo{person}{Dina
  Demner-Fushman}, \bibinfo{person}{Ellen Voorhees},
  \bibinfo{person}{William~R. Hersh}, \bibinfo{person}{Steven Bedrick},
  \bibinfo{person}{Alexander~J. Lazar}, {and} \bibinfo{person}{Shubham Pant}.}
  \bibinfo{year}{2017}\natexlab{}.
\newblock \showarticletitle{Overview of the TREC 2017 Precision Medicine
  Track}. In \bibinfo{booktitle}{\emph{TREC}}.
\newblock


\bibitem[\protect\citeauthoryear{Roberts, Demner-Fushman, Voorhees, Hersh,
  Bedrick, Lazar, Pant, and Meric-Bernstam}{Roberts et~al\mbox{.}}{2019}]%
        {pm19}
\bibfield{author}{\bibinfo{person}{Kirk Roberts}, \bibinfo{person}{Dina
  Demner-Fushman}, \bibinfo{person}{Ellen Voorhees},
  \bibinfo{person}{William~R. Hersh}, \bibinfo{person}{Steven Bedrick},
  \bibinfo{person}{Alexander~J. Lazar}, \bibinfo{person}{Shubham Pant}, {and}
  \bibinfo{person}{Funda Meric-Bernstam}.} \bibinfo{year}{2019}\natexlab{}.
\newblock \showarticletitle{Overview of the TREC 2019 Precision Medicine
  Track}. In \bibinfo{booktitle}{\emph{TREC}}.
\newblock


\bibitem[\protect\citeauthoryear{Roberts, Demner-Fushman, Voorhees, and
  Hersh}{Roberts et~al\mbox{.}}{2016}]%
        {cds2016}
\bibfield{author}{\bibinfo{person}{Kirk Roberts}, \bibinfo{person}{Dina
  Demner-Fushman}, \bibinfo{person}{Ellen~M. Voorhees}, {and}
  \bibinfo{person}{William~R. Hersh}.} \bibinfo{year}{2016}\natexlab{}.
\newblock \showarticletitle{Overview of the TREC 2016 Clinical Decision Support
  Track}. In \bibinfo{booktitle}{\emph{TREC}}.
\newblock


\bibitem[\protect\citeauthoryear{Roberts, Simpson, Voorhees, and Hersh}{Roberts
  et~al\mbox{.}}{2015}]%
        {cds2015}
\bibfield{author}{\bibinfo{person}{Kirk Roberts}, \bibinfo{person}{Matthew~S.
  Simpson}, \bibinfo{person}{Ellen Voorhees}, {and} \bibinfo{person}{William~R.
  Hersh}.} \bibinfo{year}{2015}\natexlab{}.
\newblock \showarticletitle{Overview of the TREC 2015 Clinical Decision Support
  Track}. In \bibinfo{booktitle}{\emph{TREC}}.
\newblock


\bibitem[\protect\citeauthoryear{Rogers}{Rogers}{2000a}]%
        {LDC2000T52}
\bibfield{author}{\bibinfo{person}{Willie Rogers}.}
  \bibinfo{year}{2000}\natexlab{a}.
\newblock \bibinfo{title}{TREC Mandarin LDC2000T52}.
\newblock
\newblock
\urldef\tempurl%
\url{https://catalog.ldc.upenn.edu/LDC2000T52}
\showURL{%
\tempurl}


\bibitem[\protect\citeauthoryear{Rogers}{Rogers}{2000b}]%
        {LDC2000T51}
\bibfield{author}{\bibinfo{person}{Willie Rogers}.}
  \bibinfo{year}{2000}\natexlab{b}.
\newblock \bibinfo{title}{TREC Spanish LDC2000T51}.
\newblock
\newblock
\urldef\tempurl%
\url{https://catalog.ldc.upenn.edu/LDC2000T51}
\showURL{%
\tempurl}


\bibitem[\protect\citeauthoryear{Sandhaus}{Sandhaus}{2008}]%
        {sandhaus2008nyt}
\bibfield{author}{\bibinfo{person}{Evan Sandhaus}.}
  \bibinfo{year}{2008}\natexlab{}.
\newblock \showarticletitle{The new york times annotated corpus}.
\newblock \bibinfo{journal}{\emph{Linguistic Data Consortium, Philadelphia}}
  \bibinfo{volume}{6}, \bibinfo{number}{12} (\bibinfo{year}{2008}),
  \bibinfo{pages}{e26752}.
\newblock


\bibitem[\protect\citeauthoryear{Sequiera and Lin}{Sequiera and Lin}{2017}]%
        {Sequiera2017Finally}
\bibfield{author}{\bibinfo{person}{Royal Sequiera} {and} \bibinfo{person}{Jimmy
  Lin}.} \bibinfo{year}{2017}\natexlab{}.
\newblock \showarticletitle{Finally, a Downloadable Test Collection of Tweets}.
  In \bibinfo{booktitle}{\emph{SIGIR}}.
\newblock


\bibitem[\protect\citeauthoryear{Simpson, Voorhees, and Hersh}{Simpson
  et~al\mbox{.}}{2014}]%
        {cds2014}
\bibfield{author}{\bibinfo{person}{Matthew~S. Simpson},
  \bibinfo{person}{Ellen~M. Voorhees}, {and} \bibinfo{person}{William Hersh}.}
  \bibinfo{year}{2014}\natexlab{}.
\newblock \showarticletitle{Overview of the TREC 2014 Clinical Decision Support
  Track}. In \bibinfo{booktitle}{\emph{TREC}}.
\newblock


\bibitem[\protect\citeauthoryear{Smeaton and Wilkinson}{Smeaton and
  Wilkinson}{1996}]%
        {zh5}
\bibfield{author}{\bibinfo{person}{Alan Smeaton} {and} \bibinfo{person}{Ross
  Wilkinson}.} \bibinfo{year}{1996}\natexlab{}.
\newblock \showarticletitle{Spanish and Chinese Document Retrieval in TREC-5}.
  In \bibinfo{booktitle}{\emph{TREC}}.
\newblock


\bibitem[\protect\citeauthoryear{Soboroff, Huang, and Harman}{Soboroff
  et~al\mbox{.}}{2018}]%
        {news18}
\bibfield{author}{\bibinfo{person}{Ian Soboroff}, \bibinfo{person}{Shudong
  Huang}, {and} \bibinfo{person}{Donna Harman}.}
  \bibinfo{year}{2018}\natexlab{}.
\newblock \showarticletitle{TREC 2018 News Track Overview}. In
  \bibinfo{booktitle}{\emph{TREC}}.
\newblock


\bibitem[\protect\citeauthoryear{Soboroff, Huang, and Harman}{Soboroff
  et~al\mbox{.}}{2019}]%
        {news19}
\bibfield{author}{\bibinfo{person}{Ian Soboroff}, \bibinfo{person}{Shudong
  Huang}, {and} \bibinfo{person}{Donna Harman}.}
  \bibinfo{year}{2019}\natexlab{}.
\newblock \showarticletitle{TREC 2019 News Track Overview}. In
  \bibinfo{booktitle}{\emph{TREC}}.
\newblock


\bibitem[\protect\citeauthoryear{Thakur, Reimers, Rücklé, Srivastava, and
  Gurevych}{Thakur et~al\mbox{.}}{2021}]%
        {beir}
\bibfield{author}{\bibinfo{person}{Nandan Thakur}, \bibinfo{person}{Nils
  Reimers}, \bibinfo{person}{Andreas Rücklé}, \bibinfo{person}{Abhishek
  Srivastava}, {and} \bibinfo{person}{Iryna Gurevych}.}
  \bibinfo{year}{2021}\natexlab{}.
\newblock \showarticletitle{BEIR: A Heterogenous Benchmark for Zero-shot
  Evaluation of Information Retrieval Models}.
\newblock \bibinfo{journal}{\emph{arXiv preprint arXiv:2104.08663}}
  (\bibinfo{date}{4} \bibinfo{year}{2021}).
\newblock
\urldef\tempurl%
\url{https://arxiv.org/abs/2104.08663}
\showURL{%
\tempurl}


\bibitem[\protect\citeauthoryear{Thorne, Vlachos, Christodoulopoulos, and
  Mittal}{Thorne et~al\mbox{.}}{2018}]%
        {fv}
\bibfield{author}{\bibinfo{person}{James Thorne}, \bibinfo{person}{Andreas
  Vlachos}, \bibinfo{person}{Christos Christodoulopoulos}, {and}
  \bibinfo{person}{Arpit Mittal}.} \bibinfo{year}{2018}\natexlab{}.
\newblock \showarticletitle{{FEVER}: a Large-scale Dataset for Fact Extraction
  and {VER}ification}. In \bibinfo{booktitle}{\emph{Proceedings of the 2018
  Conference of the North {A}merican Chapter of the Association for
  Computational Linguistics: Human Language Technologies, Volume 1 (Long
  Papers)}}. \bibinfo{publisher}{Association for Computational Linguistics},
  \bibinfo{address}{New Orleans, Louisiana}, \bibinfo{pages}{809--819}.
\newblock
\urldef\tempurl%
\url{https://doi.org/10.18653/v1/N18-1074}
\showDOI{\tempurl}


\bibitem[\protect\citeauthoryear{Voorhees}{Voorhees}{2004}]%
        {Voorhees2004Rb}
\bibfield{author}{\bibinfo{person}{Ellen Voorhees}.}
  \bibinfo{year}{2004}\natexlab{}.
\newblock \showarticletitle{Overview of the TREC 2004 Robust Retrieval Track}.
  In \bibinfo{booktitle}{\emph{TREC}}.
\newblock


\bibitem[\protect\citeauthoryear{Voorhees, Alam, Bedrick, Demner-Fushman,
  Hersh, Lo, Roberts, Soboroff, and Wang}{Voorhees et~al\mbox{.}}{2020}]%
        {treccovid}
\bibfield{author}{\bibinfo{person}{E. Voorhees}, \bibinfo{person}{Tasmeer
  Alam}, \bibinfo{person}{Steven Bedrick}, \bibinfo{person}{Dina
  Demner-Fushman}, \bibinfo{person}{W. Hersh}, \bibinfo{person}{Kyle Lo},
  \bibinfo{person}{Kirk Roberts}, \bibinfo{person}{I. Soboroff}, {and}
  \bibinfo{person}{Lucy~Lu Wang}.} \bibinfo{year}{2020}\natexlab{}.
\newblock \showarticletitle{TREC-COVID: Constructing a Pandemic Information
  Retrieval Test Collection}.
\newblock \bibinfo{journal}{\emph{ArXiv}}  \bibinfo{volume}{abs/2005.04474}
  (\bibinfo{year}{2020}).
\newblock


\bibitem[\protect\citeauthoryear{Voorhees}{Voorhees}{2005}]%
        {Voorhees2005Rb}
\bibfield{author}{\bibinfo{person}{Ellen~M. Voorhees}.}
  \bibinfo{year}{2005}\natexlab{}.
\newblock \showarticletitle{Overview of the TREC 2005 Robust Retrieval Track}.
  In \bibinfo{booktitle}{\emph{TREC}}.
\newblock


\bibitem[\protect\citeauthoryear{Wachsmuth, Syed, and Stein}{Wachsmuth
  et~al\mbox{.}}{2018}]%
        {ag}
\bibfield{author}{\bibinfo{person}{Henning Wachsmuth}, \bibinfo{person}{Shahbaz
  Syed}, {and} \bibinfo{person}{Benno Stein}.} \bibinfo{year}{2018}\natexlab{}.
\newblock \showarticletitle{Retrieval of the Best Counterargument without Prior
  Topic Knowledge}. In \bibinfo{booktitle}{\emph{Proceedings of the 56th Annual
  Meeting of the Association for Computational Linguistics (Volume 1: Long
  Papers)}} (Melbourne, Australia). \bibinfo{publisher}{Association for
  Computational Linguistics}, \bibinfo{pages}{241--251}.
\newblock
\urldef\tempurl%
\url{http://aclweb.org/anthology/P18-1023}
\showURL{%
\tempurl}


\bibitem[\protect\citeauthoryear{Wadden, Lin, Lo, Wang, van Zuylen, Cohan, and
  Hajishirzi}{Wadden et~al\mbox{.}}{2020}]%
        {sf}
\bibfield{author}{\bibinfo{person}{David Wadden}, \bibinfo{person}{Shanchuan
  Lin}, \bibinfo{person}{Kyle Lo}, \bibinfo{person}{Lucy~Lu Wang},
  \bibinfo{person}{Madeleine van Zuylen}, \bibinfo{person}{Arman Cohan}, {and}
  \bibinfo{person}{Hannaneh Hajishirzi}.} \bibinfo{year}{2020}\natexlab{}.
\newblock \showarticletitle{Fact or Fiction: Verifying Scientific Claims}. In
  \bibinfo{booktitle}{\emph{Proceedings of the 2020 Conference on Empirical
  Methods in Natural Language Processing (EMNLP)}}.
  \bibinfo{publisher}{Association for Computational Linguistics},
  \bibinfo{address}{Online}, \bibinfo{pages}{7534--7550}.
\newblock
\urldef\tempurl%
\url{https://doi.org/10.18653/v1/2020.emnlp-main.609}
\showDOI{\tempurl}


\bibitem[\protect\citeauthoryear{Wang, Lo, Chandrasekhar, Reas, Yang, Eide,
  Funk, Kinney, Liu, Merrill, Mooney, Murdick, Rishi, Sheehan, Shen, Stilson,
  Wade, Wang, Wilhelm, Xie, Raymond, Weld, Etzioni, and Kohlmeier}{Wang
  et~al\mbox{.}}{2020}]%
        {Wang2020CORD19TC}
\bibfield{author}{\bibinfo{person}{Lucy~Lu Wang}, \bibinfo{person}{Kyle Lo},
  \bibinfo{person}{Yoganand Chandrasekhar}, \bibinfo{person}{Russell Reas},
  \bibinfo{person}{Jiangjiang Yang}, \bibinfo{person}{Darrin Eide},
  \bibinfo{person}{K. Funk}, \bibinfo{person}{Rodney~Michael Kinney},
  \bibinfo{person}{Ziyang Liu}, \bibinfo{person}{W. Merrill},
  \bibinfo{person}{P. Mooney}, \bibinfo{person}{D. Murdick},
  \bibinfo{person}{Devvret Rishi}, \bibinfo{person}{Jerry Sheehan},
  \bibinfo{person}{Zhihong Shen}, \bibinfo{person}{B. Stilson},
  \bibinfo{person}{A. Wade}, \bibinfo{person}{K. Wang},
  \bibinfo{person}{Christopher Wilhelm}, \bibinfo{person}{Boya Xie},
  \bibinfo{person}{D. Raymond}, \bibinfo{person}{Daniel~S. Weld},
  \bibinfo{person}{Oren Etzioni}, {and} \bibinfo{person}{Sebastian Kohlmeier}.}
  \bibinfo{year}{2020}\natexlab{}.
\newblock \showarticletitle{CORD-19: The Covid-19 Open Research Dataset}.
\newblock \bibinfo{journal}{\emph{ArXiv}} (\bibinfo{year}{2020}).
\newblock


\bibitem[\protect\citeauthoryear{Wilkinson}{Wilkinson}{1997}]%
        {zh6}
\bibfield{author}{\bibinfo{person}{Ross Wilkinson}.}
  \bibinfo{year}{1997}\natexlab{}.
\newblock \showarticletitle{Chinese Document Retrieval at TREC-6}. In
  \bibinfo{booktitle}{\emph{TREC}}.
\newblock


\bibitem[\protect\citeauthoryear{Wolf, Lhoest, von Platen, Jernite, Drame, Plu,
  Chaumond, Delangue, Ma, Thakur, Patil, Davison, Scao, Sanh, Xu, Patry,
  McMillan-Major, Brandeis, Gugger, Lagunas, Debut, Funtowicz, Moi, Rush,
  Schmidd, Cistac, Muštar, Boudier, and Tordjmann}{Wolf et~al\mbox{.}}{2020}]%
        {2020HuggingFace-datasets}
\bibfield{author}{\bibinfo{person}{Thomas Wolf}, \bibinfo{person}{Quentin
  Lhoest}, \bibinfo{person}{Patrick von Platen}, \bibinfo{person}{Yacine
  Jernite}, \bibinfo{person}{Mariama Drame}, \bibinfo{person}{Julien Plu},
  \bibinfo{person}{Julien Chaumond}, \bibinfo{person}{Clement Delangue},
  \bibinfo{person}{Clara Ma}, \bibinfo{person}{Abhishek Thakur},
  \bibinfo{person}{Suraj Patil}, \bibinfo{person}{Joe Davison},
  \bibinfo{person}{Teven~Le Scao}, \bibinfo{person}{Victor Sanh},
  \bibinfo{person}{Canwen Xu}, \bibinfo{person}{Nicolas Patry},
  \bibinfo{person}{Angie McMillan-Major}, \bibinfo{person}{Simon Brandeis},
  \bibinfo{person}{Sylvain Gugger}, \bibinfo{person}{François Lagunas},
  \bibinfo{person}{Lysandre Debut}, \bibinfo{person}{Morgan Funtowicz},
  \bibinfo{person}{Anthony Moi}, \bibinfo{person}{Sasha Rush},
  \bibinfo{person}{Philipp Schmidd}, \bibinfo{person}{Pierric Cistac},
  \bibinfo{person}{Victor Muštar}, \bibinfo{person}{Jeff Boudier}, {and}
  \bibinfo{person}{Anna Tordjmann}.} \bibinfo{year}{2020}\natexlab{}.
\newblock \showarticletitle{Datasets}.
\newblock \bibinfo{journal}{\emph{GitHub. Note:
  https://github.com/huggingface/datasets}}  \bibinfo{volume}{1}
  (\bibinfo{year}{2020}).
\newblock


\bibitem[\protect\citeauthoryear{Yang, Fang, and Lin}{Yang
  et~al\mbox{.}}{2017}]%
        {Yang2017AnseriniET}
\bibfield{author}{\bibinfo{person}{Peilin Yang}, \bibinfo{person}{Hui Fang},
  {and} \bibinfo{person}{Jimmy Lin}.} \bibinfo{year}{2017}\natexlab{}.
\newblock \showarticletitle{Anserini: Enabling the Use of Lucene for
  Information Retrieval Research}.
\newblock \bibinfo{journal}{\emph{SIGIR}} (\bibinfo{year}{2017}).
\newblock


\bibitem[\protect\citeauthoryear{Yang, Qi, Zhang, Bengio, Cohen, Salakhutdinov,
  and Manning}{Yang et~al\mbox{.}}{2018}]%
        {hp}
\bibfield{author}{\bibinfo{person}{Zhilin Yang}, \bibinfo{person}{Peng Qi},
  \bibinfo{person}{Saizheng Zhang}, \bibinfo{person}{Yoshua Bengio},
  \bibinfo{person}{William Cohen}, \bibinfo{person}{Ruslan Salakhutdinov},
  {and} \bibinfo{person}{Christopher~D. Manning}.}
  \bibinfo{year}{2018}\natexlab{}.
\newblock \showarticletitle{{H}otpot{QA}: A Dataset for Diverse, Explainable
  Multi-hop Question Answering}. In \bibinfo{booktitle}{\emph{Proceedings of
  the 2018 Conference on Empirical Methods in Natural Language Processing}}.
  \bibinfo{publisher}{Association for Computational Linguistics},
  \bibinfo{address}{Brussels, Belgium}, \bibinfo{pages}{2369--2380}.
\newblock
\urldef\tempurl%
\url{https://doi.org/10.18653/v1/D18-1259}
\showDOI{\tempurl}


\bibitem[\protect\citeauthoryear{Yates, Jose, Zhang, and Lin}{Yates
  et~al\mbox{.}}{2020}]%
        {yates2020capreolus}
\bibfield{author}{\bibinfo{person}{Andrew Yates}, \bibinfo{person}{Kevin~Martin
  Jose}, \bibinfo{person}{Xinyu Zhang}, {and} \bibinfo{person}{Jimmy Lin}.}
  \bibinfo{year}{2020}\natexlab{}.
\newblock \showarticletitle{Flexible IR pipelines with Capreolus}. In
  \bibinfo{booktitle}{\emph{Proceedings of the 29th ACM International
  Conference on Information \& Knowledge Management}}.
  \bibinfo{pages}{3181--3188}.
\newblock


\bibitem[\protect\citeauthoryear{Zuccon, Palotti, Goeuriot, Kelly, Lupu,
  Pecina, M{\"u}ller, Budaher, and Deacon}{Zuccon et~al\mbox{.}}{2016}]%
        {Zuccon2016CLEF}
\bibfield{author}{\bibinfo{person}{Guido Zuccon}, \bibinfo{person}{Joao
  Palotti}, \bibinfo{person}{Lorraine Goeuriot}, \bibinfo{person}{Liadh Kelly},
  \bibinfo{person}{Mihai Lupu}, \bibinfo{person}{Pavel Pecina},
  \bibinfo{person}{Henning M{\"u}ller}, \bibinfo{person}{Julie Budaher}, {and}
  \bibinfo{person}{Anthony Deacon}.} \bibinfo{year}{2016}\natexlab{}.
\newblock \showarticletitle{The IR Task at the CLEF eHealth Evaluation Lab
  2016: User-centred Health Information Retrieval}. In
  \bibinfo{booktitle}{\emph{CLEF}}.
\newblock


\end{thebibliography}

\end{document}